\documentclass{jfm}

\usepackage{graphicx,psfrag,color}
\usepackage{epsfig,amsmath}
\usepackage{amsfonts}
\usepackage{mathrsfs}
\usepackage{amssymb}
\usepackage{bm}
\usepackage{url}
\usepackage{color}
\usepackage{natbib}
\usepackage{mdwlist}

\newcommand{\R}{\mathbb{R}}
\newcommand{\mbf}[1]{{\mathbf{#1}}}
\newcommand{\ombf}[1]{\mbox{\boldmath{$#1$}}}

\begin{document}

\title[]{Symmetry-plane model of {3D} Euler flows and mapping to regular systems to improve blowup assessment using numerical and analytical solutions}

\author{Rachel M. Mulungye, 
 Dan Lucas, 
 Miguel D. Bustamante
}

\affiliation{Complex and Adaptive Systems Laboratory, School of Mathematical Sciences, University College Dublin, Belfield, Dublin 4, Ireland}

\maketitle

\begin{abstract}
Motivated by the work on stagnation-point type exact solutions (with infinite energy) of 3D Euler fluid equations by \citep{Gibbon1999497} and the subsequent demonstration of finite-time blowup by \citep{Constantin:2000fa} we introduce a one-parameter family of models of the 3D Euler fluid equations on a 2D symmetry plane. Our models are seen as a deformation of the 3D Euler equations which respects the variational structure of the original equations so that explicit solutions can be found for the supremum norms of the basic fields: vorticity and stretching rate of vorticity. In particular, the value of the model's parameter determines whether or not there is finite-time blowup, and the singularity time can be computed explicitly in terms of the initial conditions and the model's parameter. We use a representative of this family of models, whose solution blows up at a finite time, as a benchmark for the systematic study of errors in numerical simulations. Using a high-order pseudospectral method, we compare the numerical integration of our ``original'' model equations against a ``mapped'' version of these equations. The mapped version is a globally regular (in time) system of equations, obtained via a bijective nonlinear mapping of time and fields from the original model equations. The mapping can be constructed explicitly whenever a Beale-Kato-Majda type of theorem is available therefore it is applicable to the 3D Euler equations  \citep{bustamante20113d}. We show that the mapped system's numerical solution leads to more accurate (by three orders of magnitude) estimates of supremum norms and singularity time compared to the original system. The numerical integration of the mapped equations is demonstrated to entail only a small extra computational cost. We study the Fourier spectrum of the model's numerical solution and find that the analyticity-strip width (a measure of the solution's analyticity) tends to zero as a power law in a finite time. This is in agreement with the finite-time blowup of the fields' supremum norms, in the light of rigorous bounds stemming from the bridge \citep{bustamante2012interplay} between the analyticity-strip method and the Beale-Kato-Majda type of theorems. We conclude by discussing the implications of this research on the  analysis of numerical solutions to the 3D Euler fluid equations.
\end{abstract}
\normalsize
\vspace{2pc}

\section{Introduction}

The three dimensional incompressible Euler fluid equations represent a triple point between the areas of engineering, physics and mathematics. Originally derived by Leonhard Euler \citep{Euler1752}, these equations have stood firm after 250 years of research, playing a pivotal role in the description of fluids of all types. This pivotal role lies in the mathematical modelling and numerical simulations of physical phenomena taking place in fluids. One of the main challenges these equations pose is that it is not known in detail how the energy content is transferred throughout spatial scales. Efforts towards understanding this cascade process have generated significant cross-fertilisation across disciplines of research. For real-life problems, understanding this process is needed in order to optimise industrial production of metallic alloys, gas and oil extraction and transport, and performance of turbo-machinery in general. From atmospheric science and oceanography to plasma physics, the governing fluid equations share the same feature: nonlinear terms due to advection and pressure, which carry along transfers of energy throughout different length scales making simulation and modelling a very difficult task. The main difficulties from the practical point of view have to do with accuracy and stability of the numerical solutions. For example, numerical weather prediction relies on accurate models to improve the skill of a forecast. Interestingly, the same difficulties arise in the mathematical problem of determining whether the solution of the 3D Euler equations develops a singularity in a finite time. Accurate modelling of the 3D Euler and Navier-Stokes fluid equations can also shed light on the unsolved problem of turbulence \citep{1941DoSSR..30..301K}, defined as a hypothetical out-of-equilibrium nonlinear regime characterised by intermittent fluxes of energy across scales, whereby the fluid's degrees of freedom exhibit quasi-periodic oscillations that are amenable to statistical analyses.

One of the most important conditional results known to date regarding regularity of classical (as opposed to `weak') solutions to the 3D Euler fluid equations, is the so-called Beale-Kato-Majda (BKM) theorem \citep{BKM84}, which states that all $L^2$ Sobolev norms of the velocity field are bounded up to time $T$ provided the time integral, up to time $T$, of the supremum norm of vorticity is finite. For 3D Euler fluid equations and other ideal equations, Cauchy's Lagrangian formulation (and Kelvin's circulation theorem) dictate that a loss of regularity must be accompanied by the collapse of vortex tubes, or regions of localised vorticity. Therefore, for a given simulation, numerical methods and diagnostics will have progressive difficulties in resolution and efforts must concentrate on resolving the spatial scales.

Here we attempt to review the extensive literature on the problem of finite-time blow-up in the 3D Euler equations. In the interests of brevity we direct the reader to the reviews \citep{Bardos07eulerequations, Gibbon20081894} where pre-2008 efforts are summarised and include several papers in the Proceedings of the international conference ``Euler equations: 250 years on'', notably \citep{grafke2008numerical, Bustamante:2008p2289}. As for post-2008 efforts, we highlight:

In \citep{Orlandi:2014kl}, finite difference methods are used to study numerical simulations of the 3D Euler \& Navier-Stokes equations in the context of the Chaplygin-Lamb dipole initial conditions. These initial conditions have reduced regularity in the sense that only low-order $L^2$ Sobolev norms of the velocity field exist: the initial vorticity has bounded support. The low regularity of the solution in this case makes most of the available theorems inapplicable, so the focus is put on the power-law of the energy spectrum, $E(k,t) \sim k^{-n(t)},$ where the exponent $n(t)$ seems to tend to the value $3$ at late times, consistent with a finite-time singularity in 3D Euler. 

\cite{Kerr:2013gh} returns to an extended geometry case of the well studied \citep{Kerr:1993p2227,Hou:2006p2135} antiparallel vortex tube candidate initial conditions. Kerr uses new bounds on $L^p$ norms of the vorticity field introduced by \citep{Gibbon:2013cg} and shows that the system has two distinct behaviours; an early-time power law growth of (ordered) moments and late-time super-exponential growth of moments (with broken ordering). The analysis of Gibbon's moments is extended in \citep{Donzis:2013ed} via four different pseudo-spectral methods by different research groups, to simulate 3D Euler \& Navier-Stokes equations, with regular (in fact, analytical) initial conditions, obtaining evidence against finite-time singularity in both Euler and Navier-Stokes.

In \citep{Grafke:2013dy}, adaptive mesh refinement methods are applied to the study of depletion of nonlinearity in the simulation of 3D Euler equations, with careful analyses of the bounds introduced by \citep{Deng:2005p2138} on the local behaviour of vortex-line length and curvature near the vorticity maximum, with analytical initial conditions of the Kida-Pelz type, leading to no finite-time singularity.

In \citep{Luo:2014jg}, the role of boundaries was addressed in a numerical simulation of axisymmetric 3D Euler equations in a cylinder, with strong evidence for a finite-time blowup. Boundaries are also the subject of a recent work by \cite{Kiselev:2014ti} who show that the normally regular 2D Euler equations can exhibit finite-time singularity in a norm of vorticity when non-smooth bounded domains are considered.

To our knowledge, a thorough study is yet to materialise about the role of initial conditions on the singularity of the 3D Euler or Navier-Stokes equations. However, important steps towards this understanding have been taken in terms of nonlinear optimisation of initial conditions, starting with \citep{Lu:2008dz} and notably by \citep{Ayala:2014hi} in the 2D context.

It is worth mentioning some approaches that have tackled successfully other models, related to but differing from 3D Euler in key technical aspects that allow for exact results. Arguably the first example of an integrable inviscid fluid singularity was presented by \cite{1984PhyA..125..150V} where the self motion of a Lagrangian ``free'' fluid element was considered via a local expansion of velocity and a pressure ansatz which, while satisfying conservation of angular momentum, allows energy to grow (see also further work by \cite{Cantwell:1992jp}). Finite-time singularity was demonstrated in the generalised surface quasi-geostrophic equation: see  \citep{Rodrigo2009} and references therein. In shell models of turbulence, \citep{PhysRevE.87.053011} demonstrate that inertial-range cascades of energy transfers are due to the succession of intermittent coherent structures in the form of finite-time blowups, described by universal self-similar characteristics. The Hamiltonian approach introduced by \citep{kuznetsov2000hamiltonian} allows the authors to deform the 3D Euler equations to an integrable model while keeping the vortex-line structure, establishing rigorously a finite-time blowup scenario based on the breaking of vortex lines. 

This paper introduces a new one-parameter family of models of 3D Euler on a 2D symmetry plane, motivated by the work on stagnation-point type of exact solutions (with infinite energy) of 3D Euler fluid equations by \citep{Gibbon1999497} and the subsequent demonstration of finite-time blowup by \citep{Constantin:2000fa}. Our new models are seen as a deformation of the 3D Euler equations, which  still respect the variational structure of the original equations so that explicit solutions can be found for the supremum norms of the basic fields. In particular, the value of the model's parameter determines whether there is a finite-time blowup, and the singularity time can be computed explicitly in terms of the initial conditions and the model's parameter.

The state of the art to be exploited in this paper spins out of the Beale-Kato-Majda theorem as a set of interesting applications:
\begin{enumerate}
\item[(i)] The bijective mapping to regular fields introduced by \citep{bustamante20113d}, which is a nonlinear mapping of both time and velocity field, that transforms the original system to a globally (in time) regular system. The solution of the mapped system is amenable to numerical simulations using the same methods as in the original system, and the evidence indicates that the numerical simulation of the mapped equations should provide more accurate results than the numerical simulation of the original equations. The applicability of this mapping has a wide range, including 3D and 2D models of Euler, Navier-Stokes and magneto-hydrodynamics (MHD), and we will apply it to our model.
\item[(ii)] The bridge between the BKM theorem and the analyticity-strip method, developed in \citep{bustamante2012interplay} for 3D Euler and applied in \citep{brachet2013ideal} for 3D MHD. This bridge implies that, if the initial condition is analytic with analyticity-strip width $\delta_0,$ then the local blowup of a quantity (say a supremum norm of some field) must be accompanied by a fast-enough loss of analyticity of the solution. In the case of a finite-time singularity this means that the instantaneous analyticity-strip width $\delta(t)$ must go to zero in a finite time, \emph{at a fast-enough rate}. This is applicable to our model.
\end{enumerate}

The structure of this paper is as follows. In Section \ref{sec:3DEuler} we introduce the 3D Euler fluid equations and then restrict the analysis to the so-called symmetry plane. We find that the evolution on the plane is determined by two scalar fields: vorticity and stretching rate. We obtain a rigorous system  of evolution equations for these fields and show that the equations are not closed: knowledge of the 3D flow is needed in order to get a pressure term on the plane. However, we demonstrate that a simple closure condition is sufficient in order to model this pressure term, thus generalising the condition on the pressure term by \citep{Gibbon1999497}. In this way we introduce a one-parameter family of models satisfying the closure condition.

In Section  \ref{sec:SP_ana} we provide the analytical solution along characteristics of this family of models and show explicitly that the fields have a finite-time singularity, for generic initial conditions and for some choices of the model's parameter. In particular, the singularity time is found analytically in terms of the initial condition and the value of the model's parameter.

In Section \ref{sec:mapping} we apply the method introduced by \citep{bustamante20113d} on mapping bijectively the original variables to a globally regular system, with mapped time and fields that are based on the existence of a Beale-Kato-Majda type of theorem \citep{BKM84}. The analytic and numerical advantages of working on the mapped variables are discussed. The analytical solutions for the blowup quantities are derived in terms of the mapped variables. Importantly, formulae for the original blowup quantities in terms of the mapped variables are presented. In particular, an estimate of the singularity time of the original system is obtained in terms of the mapped system's numerical solution.

Section \ref{sec:comparison} presents a comparison of the numerical solution of the original symmetry plane model and its mapped counterpart for the assessment of finite-time singularity. We monitor errors in several quantities relative to the analytic solution and discuss a number of nuances which arise. With reference to the works of \citep{Sulem:1983, bustamante2012interplay}, we present a thorough study of the spectra of stretching rate and investigate the spatial structure of the blow-up via the analyticity strip method. Finally we consider the estimation of singularity time from both systems and demonstrate a robust improvement on using the mapped system even when considering the additional computational burden it incurs.

Finally in Section \ref{sec:concl} we summarise and highlight the most important results, and discuss scope for using our methods in the research of the 3D Euler singularity problem.

\section{3D Euler fluid near a symmetry plane}
\label{sec:3DEuler}
\subsection{3D Euler fluid equations}

 Let us consider the 3D incompressible Euler equations for the velocity field
$\mbf{u}(x,y,z,t) \in \R^3$ defined for $(x,y,z) \in \R^3$ and in a time interval $t \in [0,T)$:
\begin{eqnarray}
\label{eq:Euler}
 \frac{\partial \mbf{u}}{\partial t} + \mbf{u} \cdot \nabla \mbf{u} = - \nabla p\,, \qquad \quad \nabla \cdot \mbf{u} = 0.
\end{eqnarray}
We define the vorticity field as the three-dimensional vector field $\ombf{\omega} \equiv \nabla \wedge \mbf{u}\,.$

\subsection{Symmetry plane}
We consider a special configuration of the fields and define a symmetry plane at $z = 0$ by the following conditions on the velocity and pressure fields:
\begin{equation}
\label{eq:sym_plane}
\begin{array}{cccc}
u_x(x,y,z,t) = u_x(x,y,-z,t) & u_y(x,y,z,t) = u_y(x,y,-z,t) \\
p(x,y,z,t) = p(x,y,-z,t) & u_z(x,y,z,t) = -u_z(x,y,-z,t) \,,
\end{array}
\end{equation}
for arbitrary $(x,y,z) \in \R^3$ and $t \in [0,T).$ It is easy to show that these conditions are consistent with the evolution equations (\ref{eq:Euler}). As for the vorticity, these conditions imply
\begin{equation*}
\begin{array}{rcl}
\omega_x(x,y,z,t) = -\omega_x(x,y,-z,t) & \omega_y(x,y,z,t) = -\omega_y(x,y,-z,t) & \omega_z(x,y,z,t) = \omega_z(x,y,-z,t)\,.
\end{array}
\end{equation*}
for all $(x,y,z) \in \R^3$ and $t \in [0,T).$

At the symmetry plane $z=0$, the 3D Euler fluid equations will simplify because: (i) $u_z(x,y,0,t) \equiv 0$ so the velocity field is parallel to the plane, (ii) $\omega_x(x,y,0,t) = \omega_y(x,y,0,t) = 0$ so the vorticity field is perpendicular to the plane. This leads to a new system of equations which is ``almost'' 2D (except for a pressure term depending on the full 3D velocity field). Let us denote the ``horizontal'' component of the velocity field and the pressure at the symmetry plane by
$$\mbf{u}_\mathrm{h}(x,y,t) \equiv (u_x(x,y,0,t), u_y(x,y,0,t))\,,\qquad p_\mathrm{h}(x,y,t) \equiv p(x,y,0,t)\,.$$
The horizontal components of equations (\ref{eq:Euler}) become, at $z=0$,
\begin{eqnarray}
\label{eq:Euler2D}
 \frac{\partial \mbf{u}_\mathrm{h}}{\partial t} + \mbf{u}_\mathrm{h} \cdot \nabla_\mathrm{h} \mbf{u}_\mathrm{h} = - \nabla_\mathrm{h} p_\mathrm{h}\,,
\end{eqnarray}
where $\nabla_\mathrm{h}$ denotes the ``horizontal'' gradient operator, $\nabla_\mathrm{h} = (\partial_x, \partial_y).$ The incompressibility condition in Eqs.~(\ref{eq:Euler}) allows us to define the stretching-rate scalar on the symmetry plane:
\begin{eqnarray}
\label{eq:def_gamma}
\gamma(x,y,t) \equiv u_{z,z}(x,y,0,t) = - \nabla_\mathrm{h} \cdot \mbf{u}_\mathrm{h}(x,y,t)\,.
\end{eqnarray}
Therefore, even though $u_{z} = 0$ at the symmetry plane $z=0$, we have $u_{z,z}\neq 0$ at $z=0.$ Let us compute the $z$-derivative of the $z$-component of equation (\ref{eq:Euler}) and then evaluate at $z=0$. We obtain $\frac{\partial \gamma}{\partial t} + \left.\partial_z(u_x u_{z,x} + u_y u_{z,y} + u_z u_{z,z})\right|_{z=0} = - \left.p_{, zz}\right|_{z=0}\,.$
This can be simplified by noticing that the symmetry-plane conditions (\ref{eq:sym_plane}) imply $u_z = u_{z,zz} = 0$ and $u_{x,z} = u_{y,z} = 0$ at $z=0$  so that we are left with
\begin{eqnarray}
\label{eq:Euler_uzz}
 \frac{\partial \gamma}{\partial t} + \mbf{u}_\mathrm{h} \cdot \nabla_\mathrm{h} \gamma + \gamma^2 = - \left.p_{, zz}\right|_{z=0}\,.
\end{eqnarray}
Though this equation does not close (the pressure still depends on the full 3D velocity profile), it is remarkable that the equation for the vorticity at the symmetry plane does close. As mentioned before, the vorticity at the symmetry plane has no horizontal component so we can define the vorticity scalar
\begin{eqnarray}
\label{eq:def_omega}
\omega(x,y,t) \equiv \omega_z(x,y,0,t) = \partial_x u_y - \partial_y u_x\,.
\end{eqnarray}
An evolution equation for vorticity is obtained by taking the curl of the 2D equations (\ref{eq:Euler2D}). We readily obtain
\begin{eqnarray}
\label{eq:Euler2D_omega}
\frac{\partial \omega}{\partial t} + \mbf{u}_\mathrm{h} \cdot \nabla_\mathrm{h} \omega  = \gamma \omega\,,
\end{eqnarray}
which explains the meaning of $\gamma$ as the stretching rate of vorticity.

Taken together, equations (\ref{eq:def_gamma})--(\ref{eq:Euler2D_omega}) would be a closed system in two dimensions, \emph{except for the pressure term} which, as usual, depends on the full 3D velocity profile. An important consistency condition on the pressure term is derived after integrating spatially equation (\ref{eq:Euler_uzz}) over the whole horizontal domain, and discarding boundary terms by assuming either periodic or vanishing boundary conditions on the horizontal velocity field. The condition reads
\begin{eqnarray}
\label{eq:condition_p}
\int \int \left.p_{, zz}\right|_{z=0} dx dy = - 2 \int \int (\gamma(x,y,t))^2 dx dy\,.
\end{eqnarray}

In this paper we propose a consistent family of \emph{closure} models for the pressure term, based on an exact solution by \citep{Gibbon1999497}. These models will be discussed in the next subsections.

From here on, we will assume periodic boundary conditions in the two spatial directions $(x,y)$, for the three-dimensional velocity field components $\mbf{u} = (u_x,u_y,u_z)$ and the pressure scalar $p,$ with periodicity box $[0,2\pi]\times[0,2\pi]\,.$ As for the $z$-direction, we do not assume yet any boundary condition.

\subsection{Gibbon et al.~exact solution of 3D Euler (and Navier-Stokes)}

A general class of exact solutions of 3D Euler (and Navier Stokes) was presented by \citep{Gibbon1999497}. In the case of the 3D Euler equations in the presence of a symmetry plane, this exact solution becomes
$$\mbf{u}(x,y,z,t) = (\mbf{u}_\mathrm{h}(x,y,t), z\, \gamma(x,y,t))\,,$$
where  $\mbf{u}_\mathrm{h}$ and $\gamma$ satisfy equations (\ref{eq:def_gamma})--(\ref{eq:Euler2D_omega}), along with the following condition on the pressure:
$$p_{,zz}(x,y,z,t) = f(t)$$
(a function of time only). Due to the periodicity of the horizontal domain, condition (\ref{eq:condition_p}) implies the closure $f(t) = - 2 \langle \gamma^2 \rangle,$ where
$$\langle F \rangle \equiv \frac{1}{(2\pi)^2}\int_0^{2\pi}\int_0^{2\pi} F(x,y,t)\, dx\,dy\,.$$
Correspondingly, equations (\ref{eq:def_gamma})--(\ref{eq:Euler2D_omega}) determine the fate of the full 3D flow via the knowledge of the scalars $\gamma$ and $\omega$. Remarkably, along the characteristics of the horizontal velocity field $\mbf{u}_\mathrm{h}$ the equation for stretching rate $\gamma$ is ``decoupled'' from the system and reads
\begin{eqnarray}
\label{eq:Euler_uzz_Gibbon}
\left( \frac{\partial}{\partial t} + \mbf{u}_\mathrm{h} \cdot \nabla_\mathrm{h} \right)\gamma + \gamma^2 = 2 \langle \gamma^2 \rangle
\end{eqnarray}
This allowed \citep{Constantin:2000fa} to solve for $\gamma$ along characteristics (and for vorticity $\omega$, which can be found a posteriori), proving that the stretching rate $\gamma$ would blow up in a finite time, with explicit formulae for the singularity time which confirmed the accuracy of the numerical blowup predictions in \citep{Ohkitani20003181}. We also note here for completeness the work of \cite{Gibbon:2003ix} who also proved blow-up in the axisymmetric case.

\subsection{A physically-motivated model on the symmetry plane}

The main drawback of the exact-solution ansatz $p_{,zz} = f(t)$ is that it has infinite energy (pressure goes like $z^2$).
In particular, the exact-solution ansatz is not suitable for 3D numerical simulations on fully periodic domains, where the $z$ coordinate is also periodic. One can cure this drawback by re-interpreting the ansatz as a closure on the symmetry plane: $\left.p_{, zz}\right|_{z=0} = f(t)$ and dealing with a model rather than an exact solution. Importantly, condition (\ref{eq:condition_p}) must be met at all times. However, this closure ansatz has, by definition, a spatially uniform pressure curvature $p_{,zz}|_{z=0}$ on the symmetry plane, a feature that is not observed in numerical simulations. Therefore an improved type of ansatz is required, satisfying condition (\ref{eq:condition_p}) while at the same time showing a spatial dependence of $p_{,zz}|_{z=0}$ on the symmetry plane. We propose a one-parameter family of models which achieves all this, while still keeping the structure of characteristics, so blowup can be assessed analytically. Looking at equations (\ref{eq:def_gamma})--(\ref{eq:Euler2D_omega}), the model is defined by the closure
$$\left.p_{, zz}\right|_{z=0} = - 2 \langle \gamma^2 \rangle + \lambda \left(\gamma^2 - \langle\gamma^2\rangle\right)\,,$$
where $\lambda$ is a real (free) parameter. With this closure, the family of models corresponds to the following equations on the symmetry plane:
\begin{eqnarray}
\label{GAMMA2}
\frac{\partial \gamma}{\partial t} +\mbf{u}_\mathrm{h}\cdot\nabla_\mathrm{h} \gamma &=& (2+\lambda)\langle\gamma^2\rangle - (1+\lambda)\gamma^2\,,\\
\label{OMEGA3}
\frac{\partial \omega}{\partial t} +\mbf{u}_\mathrm{h}\cdot\nabla_\mathrm{h} \omega &=& \gamma\omega\,,
\end{eqnarray}
where $(x,y)\in \mathbb{T}^2 \equiv [0,2\pi]\times[0,2\pi]$ and $\lambda \in \R.$ The case $\lambda = 0$ recovers the equations of \citep{Gibbon1999497}. The horizontal velocity field $\mbf{u}_\mathrm{h} = (u_x, u_y)$ is defined, as usual, by equations (\ref{eq:def_gamma}) and (\ref{eq:def_omega}). The evolution equations  (\ref{GAMMA2})--(\ref{OMEGA3}) are supplemented by initial conditions $\gamma(x,y,0) = \gamma_0(x,y)$ and $\omega(x,y,0)=\omega_0(x,y),$ which must have zero mean: $\langle \gamma_0 \rangle = \langle \omega_0 \rangle = 0.$

We must stress that we do not know whether this family of models could be extended to three dimensions and become a valid solution of the full 3D Euler equations; this is a matter of further investigation. What we do know is that any smooth solution of 3D Euler with discrete mirror symmetry has a 2D plane (the symmetry plane), where vorticity is a scalar satisfying equation (\ref{OMEGA3}), even in the case of finite energy. The effective unknown is the governing equation for vorticity stretching rate at the symmetry plane. We motivate our family of models by the physical interpretation of the pressure term at the symmetry plane and the introduction of a tuneable parameter, which provides a range of phenomenology and asymptotic behaviours. This approach provides a valuable framework for developing methods for assessing blow-up.

\section{Symmetry-plane model: Analytical solutions}
\label{sec:SP_ana}
Equations (\ref{GAMMA2})--(\ref{OMEGA3}) can be solved for $\gamma$ and $\omega$ along characteristics, using a classical method that will be presented in detail in a subsequent paper. The result presented below can be verified independently by direct inspection.  We consider the case $\lambda \neq -1$ (the case $\lambda=-1$ requires a limiting procedure and is omitted here for brevity). Characteristics are two-dimensional curves $(X(t),Y(t))$ defined by the system of equations
$$\frac{d X}{dt} = u_x(X(t),Y(t),t), \qquad \frac{d Y}{dt} = u_y(X(t),Y(t),t)\,.$$
Explicitly, let the characteristic have initial condition $(X(0),Y(0)) = (X_0,Y_0)\,.$ Then, in the case $\lambda \neq -1$ the solution is
\begin{eqnarray}
\label{eq:sol_gamma}
 \gamma(X(t),Y(t),t) &=&  \frac{d}{dt}\left(\ln \left[\frac{1 + (\lambda+1)\,\gamma_0(X_0,Y_0) \,S(t)}{\dot S(t)^{1/2}}\right]^{\frac{1}{\lambda+1}}\right)\,,\\
\label{eq:sol_omega}
 \omega(X(t),Y(t),t) &=& \omega_0(X_0,Y_0) \left[\frac{1 + (\lambda+1)\,\gamma_0(X_0,Y_0) \,S(t)}{\dot S(t)^{1/2}}\right]^{\frac{1}{\lambda+1}}\,,
\end{eqnarray}
where the function $S(t)$ satisfies the following ODE:
\begin{equation}
\label{eq:S_ODE}
\dot{S}(t)=\left[\frac{1}{4\pi^2}\int_0^{2\pi}\int_0^{2\pi}[1+(\lambda+1)\gamma_0(x,y)S(t)]^{\frac {-1}{\lambda+1}} \,dx\,dy\right]^{-2(\lambda+1)}\,, \quad S(0)=0\,.
\end{equation}
This ODE is obtained due to an identity satisfied by the Jacobian of the back-to-labels transformation:
\begin{equation}
\label{eq:J}
 J(t;X_0,Y_0) = \det\left(\frac{\partial(X(t),Y(t))}{\partial(X_0,Y_0)}\right)\,.
\end{equation}
From the fact that $\nabla_\mathrm{h} \cdot \mbf{u}_\mathrm{h} = -\gamma$ we readily obtain an evolution equation for $J$ which can be solved:
\begin{equation*}
\frac{ \dot J(t;X_0,Y_0)}{J(t;X_0,Y_0)} = -  \gamma(X(t),Y(t),t) \quad \Longrightarrow \quad J(t;X_0,Y_0) = e^{-\int_0^t \gamma(X(s),Y(s),s)ds}\,,
\end{equation*}
and using equation (\ref{eq:sol_gamma}) we obtain the solution
\begin{equation}
\nonumber
 J(t;X_0,Y_0) = \left[\frac{1 + (\lambda+1)\,\gamma_0(X_0,Y_0) \,S(t)}{\dot S(t)^{1/2}}\right]^{-\frac{1}{\lambda+1}}\,.
\end{equation}

This Jacobian satisfies the identity
$(\int_{\mathbb{T}^2}dxdy =) 4\pi^2 = \int_{\mathbb{T}^2} J(t;X_0,Y_0)dX_0 dY_0\,,$
which leads to the ODE (\ref{eq:S_ODE}) satisfied by $S(t).$

Notice that Kelvin's theorem on circulation conservation, or more accurately Cauchy's invariants \citep{kuznetsov2006vortex, Frisch:2014gl}, follow directly from the fact that $J(t;X_0,Y_0) \omega(X(t),Y(t),t) = \omega_0(X_0,Y_0)$ for any characteristic's initial condition $(X_0,Y_0).$

\subsection{Blowup solutions}

The solutions along characteristics for stretching rate (\ref{eq:sol_gamma}) and vorticity (\ref{eq:sol_omega}) will develop a singularity if the factor $1 + (\lambda+1)\,\gamma_0(X_0,Y_0) \,S(t)$ becomes zero for some time $t$ and position $(X_0,Y_0).$ Since $S(0)=0$ and $\dot{S}(t) \geq 0,$ it follows that $S(t)$ can only grow in time and thus the singularity will occur first at the characteristic starting at the position of the infimum (if $\lambda > -1$) or the supremum (if $\lambda < -1$) of $\gamma_0$ over $\mathbb{T}^2.$ Consequently, the singularity time $T^*$ is defined by the condition $S(T^*)=S^*,$
where $S^*(>0)$ is defined by
$$ -\frac{1}{S^*}=\left\{
\begin{array}{rcl}
(\lambda+1){\displaystyle \sup_{(x,y)\in \mathbb{T}^2}\gamma_0(x,y)},\,\, \lambda<-1\,,\\
(\lambda+1){\displaystyle \inf_{(x,y)\in \mathbb{T}^2}\gamma_0(x,y)},\,\,\lambda>-1\,.
\end{array}\right.$$
From the ODE (\ref{eq:S_ODE}) an explicit formula for the singularity time $T^*$ is derived:
$$T^*=\int_0^{S^*}\left[\frac{1}{4\pi^2}\int_0^{2\pi}\int_0^{2\pi}[1+(\lambda+1)\gamma_0(x,y)s]^{\frac {-1}{\lambda+1}}\,dx\,dy\right]^{2(\lambda+1)} ds.$$

\begin{figure}
\centering
\includegraphics[width=0.7\textwidth]{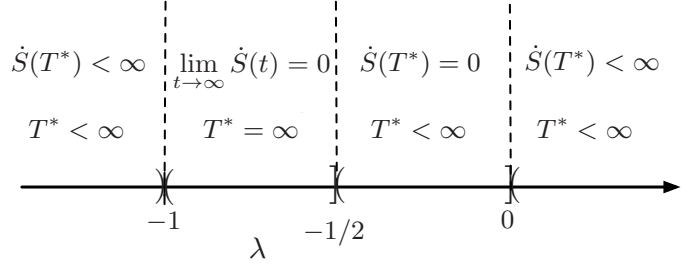}
\caption{\label{Regions} Singular/non singular behaviour of system (\ref{eq:sol_gamma})--(\ref{eq:S_ODE}) depending on the value of the model's parameter $\lambda.$ The limiting case $\lambda=-1$ (not analysed in this paper) gives an infinite-time singularity.}
\end{figure}

We consider briefly the blowup structure for the stretching and vorticity solutions for different values of the model's free parameter $\lambda$ (a detailed explanation will be presented in a forthcoming paper). The parameter space is divided in regions of finite-time blowup alternating with regions of infinite-time blowup as illustrated in figure \ref{Regions}. The region of $\lambda$ where $ \dot{S}(t)=0$ depends on the initial conditions; if the local profile of initial stretching near the infimum is parabolloidal (a generic situation), then the regions are as depicted as in figure \ref{Regions}.

\subsection{Motivation for the choice of parameter $\lambda=-3/2$ and generic initial conditions}

In this paper we will focus on one particular choice of parameter: $\lambda = -3/2.$ This choice has several advantages: 

\begin{itemize}
\item In this case the evolution equation for $S(t)$ is of the form 
$$\dot{S} = 1 + a^2 S^2\,, \qquad S(0)=0, $$
which can be solved analytically for any initial condition ($a$ depends on the initial condition), leading to explicit analytical solutions in terms of trigonometric functions for all the blowup quantities. The utility of this is that we can perform direct comparisons between theory and numerics.

\item This case gives finite-time singularity with $\dot{S}(T^*) < \infty,$ leading to simple asymptotic expressions for the blowup quantities. The singularity is controlled by the supremum of $\gamma_0$ and leads to a blowup of the form $\sup \gamma(x,y,t) \sim 2 (T^*-t)^{-1}$ when $t$ is close to the singularity time $T^*.$ This feature is analogous to what is normally expected in a 3D Euler fluid simulation in a potential singularity scenario.

\item In this case there exists a special conserved quantity: $\langle \gamma^2\rangle,$ which is reminiscent of the ``energy'' in 2D and 3D ideal models, and provides an opportunity for a simplified analysis of the Fourier spectrum.
\end{itemize}

We will exploit analytical solutions (\ref{eq:sol_omega}) for vorticity and (\ref{eq:sol_gamma}) for stretching rate in order to validate direct numerical simulations of the system. For example, using the fact that the back-to-labels transformation is bijective for $t<T^*,$ one can use equations (\ref{eq:sol_gamma})--(\ref{eq:S_ODE}) to calculate the infimum and supremum of stretching rate and vorticity from the knowledge of the initial conditions. This gives either explicit expressions in terms of simple fuctions or numerically computable expressions to any desired accuracy. In Table \ref{Cases}, we summarise  the relevant analytical solutions for the initial condition $u_x(x,y,0)=\cos(x)\sin(y)$, $u_y(x,y,0)=\cos(x)+\sin(y)$ or, in terms of stretching rate and vorticity,
\begin{eqnarray}
\label{eq:icgamma}
\gamma_0(x,y) &=& \sin(x)\sin(y)-\cos(y)\,,\\
\label{eq:icomega}
\omega_0(x,y) &=& -\sin(x)-\cos(x)\cos(y)\,.
\end{eqnarray}
Figure \ref{fig:2Dplots} shows contour plots of the initial condition for both the vorticity and  stretching rate. The discrete symmetry $(x \to \pi+x, y \to -y, u_x \to u_x, u_y \to -u_y)$ of this initial condition is preserved under the time evolution. Thus, we can restrict the analysis to the quadrant $\pi\leq x \leq2\pi, \quad \pi \leq y \leq 2\pi.$

\begin{table}
\begin{center}
    \begin{tabular}{ @{} rcl @{}}
   Case && $\lambda=-3/2$ \\[5pt]
   Initial condition  && $\gamma_0(x,y) = \sin(x)\sin(y)-\cos(y)$ \\[5pt]
      && $\omega_0(x,y) = -\sin(x)-\cos(x)\cos(y)$ \\[5pt]
   Singularity time  && $T^*=\frac{4}{\sqrt{3}}\arctan\left(\frac{\sqrt{6}}{4}\right) \approx 1.26894$ \\[5pt]
   Solution for $S(t)$ && $S(t) = \frac{4}{\sqrt{3}} {\tan \left(\frac{\sqrt{3} t}{4}\right)}$\\[5pt]
   $S(T^*) (= S^*)$ and $\dot{S}(T^*)$ && $S^* = \sqrt{2},$ \quad $\dot{S}(T^*) = 11/8$\\[5pt]
   $\left(\|\gamma(\cdot,t)\|_\infty = \right) {\displaystyle \sup_{(x,y)\in \mathbb{T}^2}\gamma(x,y,t)}=$ && $\frac{\sqrt{3}}{2}\tan\left(\frac{\sqrt{3}}{4}t\right)+\frac{\sqrt{2}}{\cos^2\left(\frac{\sqrt{3}}{4}t\right)} \frac{1}{\left(1-\frac{4}{\sqrt{6}}\tan\left(\frac{\sqrt{3}}{4}t\right)\right)}$\\[10pt]
   ${\displaystyle \inf_{(x,y)\in \mathbb{T}^2}\gamma(x,y,t)}=$ && $   \frac{11}{2 \sqrt{3} \cot \left(\frac{\sqrt{3} t}{4}\right)+4 \sqrt{2}}-\sqrt{2}
$ \quad (bounded)\\[10pt]
   $\left\langle[\gamma(\cdot,t)]^2\right\rangle_{\mathbb{T}^2}=$ && $   \frac{3}{4}
$ \quad (constant, only in the case $\lambda = -3/2$)\\[5pt]
   Vorticity at position of $\|\gamma(\cdot,t)\|_\infty$ && $\omega(\mbf{X}_\gamma(t),t) = \frac{\sec ^2\left(\frac{\sqrt{3} t}{4}\right)}{\left(1-2 \sqrt{\frac{2}{3}} \tan \left(\frac{\sqrt{3} t}{4}\right)\right)^2}$\\[5pt]
   Asymptotics as $t \to T^*$ && $\sup \gamma \sim {2}\,(T^*-t)^{-1}$\\[5pt]
    && $\sup \omega \sim \frac{16}{11}\,(T^*-t)^{-2}$\\[5pt]
     \end{tabular}
\end{center}
       \caption{Summary of analytical results for the case studied in this paper.  Supremum of vorticity is computable numerically from formula (\ref{eq:sol_omega}).
    \label{Cases}
}
\end{table}

\section{Mapping to regular fields and their evolution equations}
\label{sec:mapping}
Regardless of the availability of an analytical solution for the relevant fields, \citep{bustamante20113d} developed a general theory to study nonlinear evolution equations whose solutions present evidence of possible finite-time singularity. The main idea is to transform the original physical variables (such as velocity field) into new, so-called \emph{mapped} variables which are regular (globally in time), and thus more amenable to new analytical studies and more accurate numerical studies. The transformation, or \emph{mapping}, has applications in a wide variety of PDE models including 3D Euler/Navier-Stokes fluid equations, 3D and 2D magneto-hydrodynamics equations, Burgers equations, etc. The key ingredient to construct this mapping is a type of Beale-Kato-Majda (BKM) theorem \citep{BKM84}, which states that all relevant norms of the velocity field are bounded if and only if $\tau(t) = \int_0^t F[\mbf{u}](t') dt' \,,$ is bounded: where $F[\mbf{u}]$ is a given functional of the velocity field $\mbf{u}.$ In the case of our 2D symmetry-plane model (\ref{GAMMA2})--(\ref{OMEGA3}), following a classical analysis analogous to that in \citep{0951-7715-14-5-316} we deduce that the functional is $F[\mbf{u}_h](t') \equiv \|\gamma(\cdot,t')\|_{\infty}\,,$ the $L^{\infty}$ norm of the stretching rate $\gamma(x,y,t)$ over the spatial domain $\mathbb{T}^2.$ Explicitly, the boundedness of the integral
\begin{equation}
\label{eq:map_tau}
\tau(t) = \int_0^t \|\gamma(\cdot,t')\|_{\infty} dt'
\end{equation}
will ensure the continuity of the velocity field $\mbf{u}_h$ until time $t$ (provided the initial conditions are smooth). For example, if $\tau(t)$ is bounded then vorticity is bounded because, from equation (\ref{eq:sol_omega}), it follows
\begin{eqnarray*}
\left|\omega(X(t),Y(t),t)\right| &=& \left|\omega_0(X_0,Y_0)\right| \exp \left(\int_0^t \gamma(X(t'),Y(t'),t')dt'\right) \\
                                &\leq& \left|\omega_0(X_0,Y_0)\right| \exp \left(\int_0^t \|\gamma(\cdot,t')\|_{\infty} dt'\right) =  \left|\omega_0(X_0,Y_0)\right|  \exp(\tau(t))\,.
\end{eqnarray*}
The bijective mapping from ``original variables'' to ``mapped, regular variables'' consists of the time mapping $t \to \tau,$ equation (\ref{eq:map_tau}), along with a re-scaling of stretching rate, vorticity and velocity vector fields:
\begin{eqnarray}
\label{eq:map_gamma}
\gamma_{\mathrm{map}}(x,y,\tau) &=& \frac{\gamma (x,y,t)}{\|\gamma(\cdot,t)\|_\infty}\,,\\
\label{eq:map_omega}
\omega_{\mathrm{map}}(x,y,\tau) &=& \frac{\omega (x,y,t)}{\|\gamma(\cdot,t)\|_\infty}\,,\\
\nonumber
\Longrightarrow \mbf{u}_{\mathrm{map}}(x,y,\tau) &=& \frac{\mbf{u}_h(x,y,t)}{\|\gamma(\cdot,t)\|_\infty}\,.
\end{eqnarray}
For this bijective mapping to lead to tractable evolution equations for the mapped variables, the ``BKM'' functional $\|\gamma(\cdot,t)\|_{\infty}$ must have a time derivative that can be expressed in terms of the original variables. In our case, equation (\ref{GAMMA2}) implies
\begin{equation}
\label{eq:ODE_gamma}
 \frac{\mathrm{d}}{\mathrm{d}t}\left({\|\gamma(\cdot,t)\|_\infty}\right)= \sigma_\infty \left[(2+\lambda)\langle\gamma^2\rangle - (1+\lambda)\|\gamma(\cdot,t)\|_\infty^2\right]\,,
\end{equation}
where 
\begin{equation}
\label{eq:sigma}
\sigma_\infty \equiv \mathrm{sign} \, \gamma(\mbf{X}_{\gamma}(t),t)
\end{equation}
 is the sign of $\gamma$ at the position $\mbf{X}_{\gamma}(t)$ where the maximum of $|\gamma(\mbf{x},t)|$ is attained.

With these ingredients, the mapped variables satisfy the following system of evolution equations:
\begin{eqnarray}
\label{PDE1}
\frac{\partial\gamma_{\mathrm{map}}}{\partial\tau}+\mbf{u}_{\mathrm{map}}\cdot\nabla \gamma_{\mathrm{map}} &=&(2+\lambda)\langle\gamma_{\mathrm{map}}^2\rangle - (1+\lambda)\gamma_{\mathrm{map}}^2\nonumber \\
&& \hspace{24pt} + \sigma_\infty \,\gamma_{\mathrm{map}}\left\{1+\lambda - (2+\lambda)\langle\gamma_{\mathrm{map}}^2\rangle \right \}\\
\label{PDE2}
\frac{\partial\omega_{\mathrm{map}}}{\partial\tau} +\mbf{u}_{\mathrm{map}}\cdot\nabla \omega_{\mathrm{map}} &=& \gamma_{\mathrm{map}}\,\omega_{\mathrm{map}}+\sigma_\infty \,\omega_{\mathrm{map}}\left\{1+\lambda - (2+\lambda)\langle\gamma_{\mathrm{map}}^2\rangle \right\}.
\end{eqnarray}
These equations are supplemented with the initial constraint $\|\gamma_{\mathrm{map}}(\cdot,0)\|_{\infty} = 1.$ They differ from the original system simply by extra ``drag'' terms, which ensure that $\|\gamma_{\mathrm{map}}(\cdot,\tau)\|_{\infty} = 1$ for all $\tau < \infty.$ The most striking result is that, as the condition $\tau<\infty$ implies boundedness of the original fields, the solution of the mapped system (\ref{PDE1})--(\ref{PDE2}) is globally regular in time $\tau.$

\subsection{Using the mapped system to assess blowup of original system}
The mapping (\ref{eq:map_tau})--(\ref{eq:map_omega}) is bijective as long as $\tau<\infty.$ Correspondingly, integration of the mapped system (\ref{PDE1})--(\ref{PDE2}) should give enough information to assess blowup quantities in the original variables.\\

The norm $\|\gamma(\cdot,t)\|_{\infty}$ satisfies the ODE (\ref{eq:ODE_gamma}). In terms of $\tau$ and the mapped stretching rate $\gamma_{\mathrm{map}}(x,y,\tau),$ this equation reads
$$ \frac{\mathrm{d}}{\mathrm{d}\tau}\ln\left({\|\gamma(\cdot,t(\tau))\|_\infty}\right) = \sigma_\infty \left[(2+\lambda)\langle\gamma_{\mathrm{map}}^2\rangle - (1+\lambda)\right]\,,$$
where we have used $\frac{d\tau}{dt} = \|\gamma(\cdot,t(\tau))\|_\infty\,.$ Correspondingly we obtain, after a simple $\tau$ integration,
\begin{equation}
\label{eq:norm_gamma}
||\gamma(\cdot,t(\tau))||_\infty =\|\gamma_0\|_\infty \exp\left[-(1+\lambda)\int_0^\tau\sigma_\infty~\mathrm{d}\tau' + (2+\lambda)\int_0^{\tau}\sigma_\infty \langle\gamma_{\mathrm{map}}^2\rangle \mathrm{d}\tau'\right].
\end{equation}
Notice that the RHS is written entirely in terms of mapped fields. The integrands $\sigma_\infty$ and $\sigma_\infty\langle\gamma_{\mathrm{map}}^2\rangle$ are bounded by $1$ so, remarkably, the blowup assessment of the original variables is done in terms of bounded quantities. In particular, this leads to the following general formula for the singularity time $T^*$:

\begin{equation}
\label{eq:T_star}
T^* = \frac{1}{\|\gamma_0\|_\infty} \int_0^\infty\exp\left[(1+\lambda)\int_0^\tau\sigma_\infty~\mathrm{d}\tau' - (2+\lambda)\int_0^{\tau}\sigma_\infty \langle\gamma_{\mathrm{map}}^2\rangle \mathrm{d}\tau'\right]\mathrm{d}\tau\,.
\end{equation}
Notice that this integral converges if and only if the original problem has a finite-time singularity.\\

From equation (\ref{OMEGA3}) it follows that the norm $\|\omega(\cdot,t)\|_{\infty}$ satisfies
$$\frac{d \ln\|\omega(\cdot,t)\|_{\infty}}{dt} = \gamma(\mbf{X}_\omega(t),t)\,,$$
where $\mbf{X}_\omega(t)$ is the position at which the maximum of $|\omega(\mbf{x},t)|$ is attained. In terms of the mapped variables, this reads
$\frac{d \ln\|\omega(\cdot,t(\tau))\|_{\infty}}{d \tau} = \gamma_{\mathrm{map}}(\mbf{X}_\omega(t(\tau)),\tau)\,,$ which gives
\begin{equation}
\label{eq:norm_omega}
||\omega(\cdot,t(\tau))||_\infty =\|\omega_0\|_\infty \exp\left[\int_0^{\tau}\gamma_{\mathrm{map}}(\mbf{X}_\omega(t(\tau')),\tau') \mathrm{d}\tau'\right].
\end{equation}
Remarkably again, the blowup assessment of the original vorticity depends on a bounded quantity ($-1 \leq \gamma_{\mathrm{map}} \leq 1$).

\subsection{Analytical solution for the mapped variables in the case $\lambda=-3/2$}
The analytical solution along characteristics (\ref{eq:sol_gamma})--(\ref{eq:S_ODE}) leads to a connection with the mapped variables. Let us consider the case $\lambda=-3/2$ and the initial conditions (\ref{eq:icgamma})--(\ref{eq:icomega}). Since $\lambda<-1,$ the supremum of $\gamma$ blows up and this happens along the characteristic starting at the position of the supremum of $\gamma_0$, $\mbf{X}_+ = \left(\frac{3 \pi }{2},\frac{5 \pi }{4}\right)$ (see figure \ref{fig:2Dplots} for reference). On the other hand, the infimum of $\gamma$ (a negative quantity) remains small in size. Thus, in this case the norm $\|\gamma(\cdot,t)\|_{\infty}$ can be identified with $\sup \gamma(\cdot,t)$ (i.e. one can set $\sigma_\infty \equiv 1$) for all times $0 \leq t <T^*.$ Equation (\ref{eq:sol_gamma}) evaluated at $\mbf{X}_0 = \mbf{X}_+$ is then compared with definition (\ref{eq:map_tau}) of mapped time $\tau$ to give
\begin{equation*}
\tau(t) = \ln \frac{1}{\left[\cos \left(\frac{\sqrt{3}\, t}{4}\right)-2 \sqrt{\frac{2}{3}} \sin \left(\frac{\sqrt{3}\, t}{4}\right)\right]^2}\,.
\end{equation*}
The above relation can be inverted to solve for $t$ as a function of $\tau$, provided $t<T^*.$ The supremum of original stretching rate, in terms of $\tau,$ can be obtained after using this inversion along with the formula in Table \ref{Cases}, giving
\begin{equation}\label{gam_an_map}
\|\gamma(\cdot,t(\tau))\|_\infty = \frac{1}{2} e^{\tau /2} \sqrt{11-3 e^{-\tau }}\,, \quad 0 \leq \tau < \infty\,.
\end{equation}
There is a simple analytical expression (in terms of mapped time $\tau$) for vorticity at the position where the maximum of $|\gamma(\mbf{x},t)|$ is attained, $\mbf{X}_\gamma(t).$ Equation (\ref{eq:sol_omega}) evaluated at $\mbf{X}_0 = \mbf{X}_+$ gives
\begin{equation}\label{om_an_map}
\omega(\mbf{X}_\gamma(t(\tau)),t(\tau)) = e^{\tau}\,, \quad 0 \leq \tau < \infty\,.
\end{equation}
Notice that this is a lower bound for the $L^\infty$ norm of the vorticity. The latter norm can be obtained at all times in terms of the initial conditions, by maximising the RHS of equation (\ref{eq:sol_omega}) over the quadrant $\pi \leq x, y \leq 2\pi.$ Although this rarely leads to an explicit analytical expression for $\|\omega(\cdot,t(\tau))\|_\infty,$ it can always be computed numerically to any desired accuracy.


\section{Numerical solution of original and mapped systems and comparison with analytic solution}
\label{sec:comparison}
\subsection{Numerical solutions of the original and mapped systems}\label{sec:model}

We turn our attention to the original evolution equations (\ref{GAMMA2})--(\ref{OMEGA3}) and the mapped evolution equations (\ref{PDE1})--({\ref{PDE2}}). Both systems were solved numerically using a standard pseudospectral method written in CUDA, using CUFFT library and implemented on NVIDIA GPUs. To remove the usual aliasing errors, Hou's high-order exponential filter \citep{Hou:2007} is used, where we multiply the spectrum at each time step by the factor $\exp\left(36 (2k/N)^{36}\right),$ where $k$ is the modulus of the wavevector and $N$ is the spatial resolution. We checked that the 2/3 dealiasing rule (in which the last 1/3 of the high frequency modes are set to zero) gives similar results, but Hou's filter provides sensible spectra for a slightly broader range of wavevectors. 

In both systems time marching was carried out using the fourth order Runge-Kutta method. In the mapped system, a uniform time step $d\tau$ is used since $\tau$ stretches the temporal domain such that the singularity is at $\tau=\infty.$ This uniform $d\tau$ would imply that in the original system the time step $dt$ should be adaptive, via $dt = d\tau/\|\gamma(\cdot,t)\|_{\infty}.$  This adaptive method is normally used in 3D Euler blow-up assessment studies for reasons of accuracy close to singularity. In the results below adaptive time stepping in original variables is used for this reason, with the added advantage that the data from the original and the mapped systems are more comparably spaced. A striking result of this paper, to be shown and discussed thoroughly below in this Section, is that even within this fair-comparison scenario, the mapped system gains a significant amount of accuracy in the estimation of blowup quantities (cf. Figs. \ref{ERROR2D5}, \ref{fig:Tstar_running} and \ref{cpu_error}).

Finally,  unlike the integration of the original system, the numerical integration of the mapped system requires a special method: a normalisation so that the mapped system satisfies the constraint $||\gamma_{\mathrm{map}}(\cdot,\tau)||_\infty=1 $ , for all $\tau$. 
The accuracy of this normalisation is essential for the performance of the mapped system's numerical integration. To apply the normalisation, $||\gamma_{\mathrm{map}}(\cdot,\tau)||_\infty $ is computed using a $P_{6,k}$ quarter-section interpolation. This $P_{6,k}$ interpolation is an efficient procedure to compute the field's maximum and its position using an iterative application of cubic-splines at progressively finer resolution. This was tested against several other interpolation methods. See the discussion in appendix \ref{sect:interp}.

\begin{figure}
\centering
\vspace{12mm}
\scalebox{.8}{\input{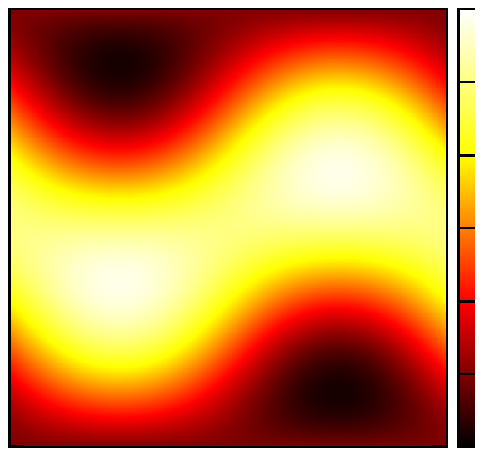}}
\vspace{15mm}
\hspace{15mm}
\scalebox{.8}{\input{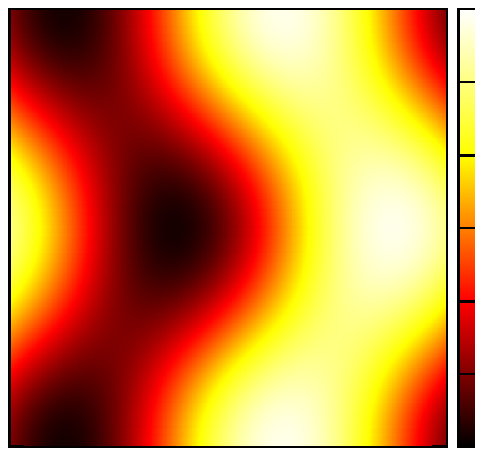}}
\vspace{9mm}
\scalebox{.8}{\input{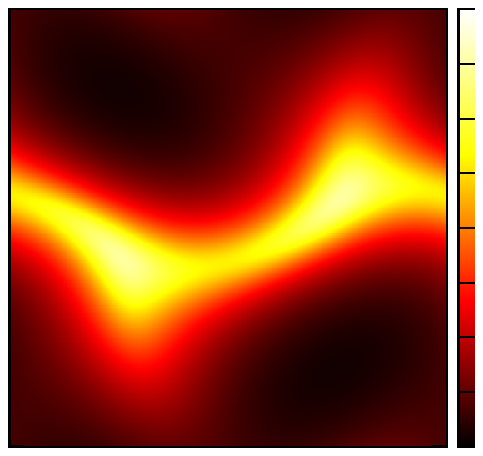}}
\vspace{6mm}
\hspace{15mm}
\scalebox{.8}{\input{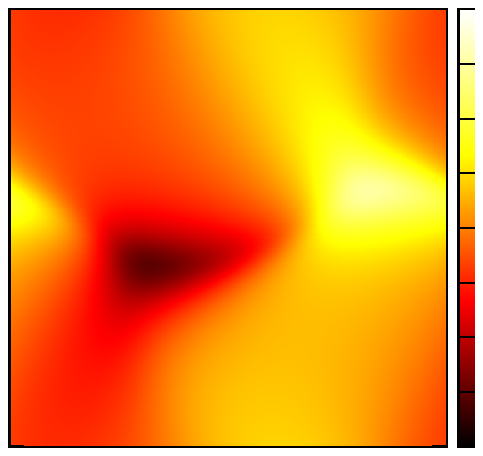}}
\vspace{4mm}
\scalebox{.8}{\input{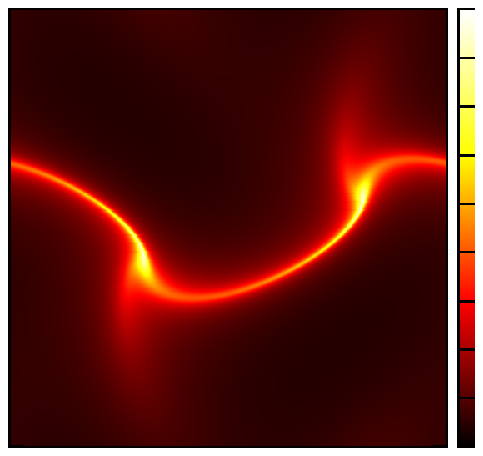}}
\vspace{2mm}
\hspace{15mm}
\scalebox{.8}{\input{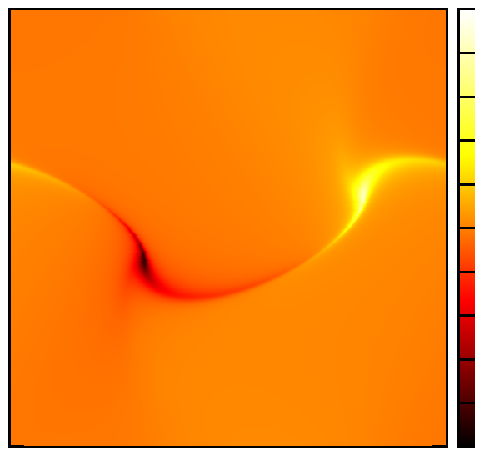}}
    \caption{Stretching rate ($\gamma$) and vorticity ($\omega$) plots for $\lambda=-3/2$ at $t=0,$ 0.5 and 1.0 (top to bottom) from the original 2D Euler model.}
   \label{fig:2Dplots}
\end{figure}

The initial conditions were chosen as in equations (\ref{eq:icgamma})--(\ref{eq:icomega}). Fig.\ref{fig:2Dplots} shows  contour plots for both the vorticity and stretching rate at various times.

\subsection{Errors in local quantities near blowup: $\|\gamma(\cdot,t)\|_{\infty}$ and $\omega(\bm{X}_\gamma(t),t)$}

A sensible definition of the error of the numerical simulation of the original 2D Euler model is the \emph{relative difference} between the supremum norm of $\gamma$  obtained from the numerical simulation and the exact analytical formula, given in Table \ref{Cases}. We can also define the error in $\omega$ by evaluating it at $\bm{X}_\gamma(t),$ the location of the supremum of $|\gamma|$, also given analytically in Table \ref{Cases}.  We define the relative errors as
\begin{align}
\mathcal{E}_\gamma(t) &= \left | \frac{\|\gamma_{\mathrm{num}}(\cdot,t)\|_{\infty}}{\|\gamma_{\mathrm{ana}}(\cdot,t)\|_{\infty}} - 1\right |,\label{errG} \\
\mathcal{E}_\omega(t) &= \left | \frac{\omega_{\mathrm{num}}(\bm{X}_\gamma(t),t)}{\omega_{\mathrm{ana}}(\bm{X}_\gamma(t),t)} - 1\right |, \label{errG}
\end{align}
where the subscripts ``num'' and ``ana'' stand for ``numerical'' and ``analytic''. While these are useful for instantaneous monitoring purposes, given a time series of a numerical solution of $\gamma$ or $\omega$ we would like to measure the error using a single number for each variable. The naive measure in terms of the $L^2$ norm of the relative error, $\|\mathcal{E}\|_2 \equiv \sqrt{\int_0^T[\mathcal{E}(t)]^2 \mathrm{d}t},$ is not the best choice because it is not bounded \emph{a priori}. Given two signals $f(t)$ and $g(t)$ for comparison, we define an $L^2$ norm of the error (\emph{not} relative error) normalised with the sum of norms of the individual signals \citep{Perlin2014}:
\[ Q(f,g) = \frac{\| f - g \|_2}{\|f\|_2+\|g\|_2} \]
which has 3 advantageous properties:\\

\begin{itemize}
\item $Q$ and $\|\mathcal{E}\|_2$ are proportional if they are small:  $\|\mathcal{E}\|_2 \propto \sqrt{T} \, Q$ if $\mathcal{E}\ll1\,.$
\item $Q$, being dimensionless, does not explicitly require a time scale ($T$) so it can be applied in a variety of contexts. In practice, though, and for assessment purposes, we will normally plot $Q$ as a function of the total integration time $T$.
\item $Q$ is bounded:  $0\leq Q \leq 1,$ with value $0$ representing perfect match and value $1$ representing  perfect mismatch.
\end{itemize}

\noindent Thus we work with $Q_\gamma = Q(\|\gamma_{\mathrm{num}}(\cdot,t)\|_{\infty},\|\gamma_{\mathrm{ana}}(\cdot,t)\|_{\infty})$ and $Q_\omega = Q(\omega_{\mathrm{num}}(\bm{X}_\gamma,t),\omega_{\mathrm{ana}}(\bm{X}_\gamma,t))$.

To gain an appropriately accurate estimate of these errors we must consider the best method for approximating $\|\gamma_{\mathrm{num}}(\cdot,t)\|_{\infty}, \bm{X}_\gamma$ and $\omega_{\mathrm{num}}(\bm{X}_\gamma,t)$. The simplest approach is to apply the maximum value across the collocation points of the discretised field, however this leads to significant spurious oscillations. The more accurate procedure is to perform some post-processing interpolation. We use the same highly-accurate interpolation as in the normalisation procedure in the mapped system. The various interpolation options are described in appendix \ref{sect:interp}.  

\begin{figure}
\centering
\scalebox{.5}{\input{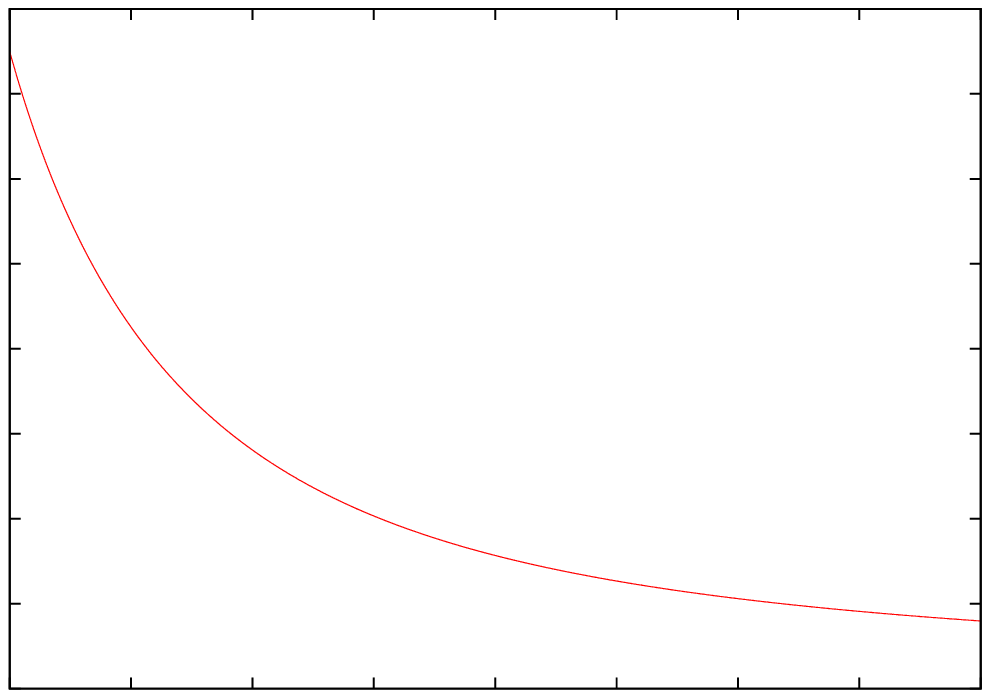}}
\scalebox{.5}{\input{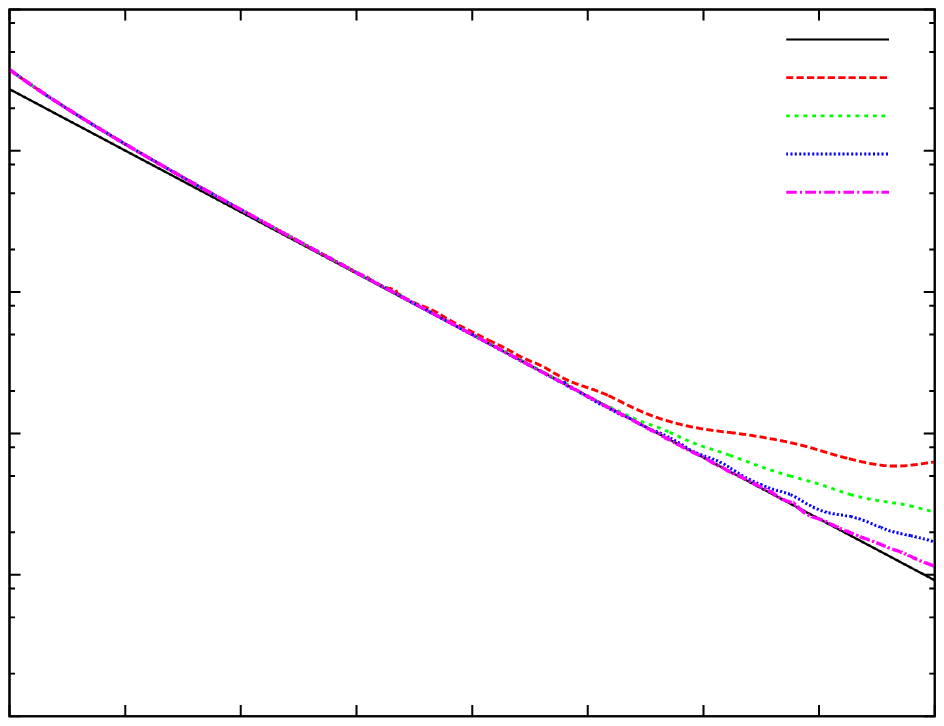}}
\caption{Left panel: Ratio between the terms $(2+\lambda)\int_0^{\tau}\sigma_\infty \langle\gamma_{\mathrm{map}}^2\rangle \mathrm{d}\tau'$ and  $-(1+\lambda)\int_0^\tau\sigma_\infty~\mathrm{d}\tau',$  case $\lambda = -3/2,$ appearing in the exponent in equation (\ref{eq:norm_gamma}). Right panel: Resolution study. Plots of $\langle\gamma_{\mathrm{map}}(\cdot, \tau)^2\rangle$ as a function of $\tau$ using data from the numerical integration of the mapped system, at different resolutions: from top to bottom, $N=256, 512,1024, 2048.$ The curves converge to the analytically-obtained asymptotic regime $\langle\gamma_{\mathrm{map}}(\cdot, \tau)^2\rangle \sim (3/11) \exp(-\tau).$   \label{fig:ratio}
}
\end{figure}

The numerical solution of the mapped system does not provide direct access to the original variable $||\gamma(\cdot,t)||_\infty $ so in order to compute it we employ equation (\ref{eq:norm_gamma}), namely $\|\gamma(\cdot,t(\tau))\|_\infty =\|\gamma_0\|_\infty \exp\left[-(1+\lambda)\int_0^\tau\sigma_\infty~\mathrm{d}\tau' + (2+\lambda)\int_0^{\tau}\sigma_\infty \langle\gamma_{\mathrm{map}}^2\rangle \mathrm{d}\tau'\right],$ where $\int_0^{\tau}\sigma_\infty\langle\gamma_{\mathrm{map}}^2\rangle \mathrm{d}\tau$ is computed using Simpson's rule. We compare this against equation (\ref{gam_an_map}). Notice that we are using the global quantity $\langle\gamma_{\mathrm{map}}^2\rangle$ for the assessment of the local quantity $\|\gamma(\cdot,t)\|_{\infty}.$ Therefore, errors of the global quantity $\langle\gamma_{\mathrm{map}}^2\rangle$  might affect the errors of this assessment. 
To see at which times these errors might be important, figure \ref{fig:ratio} (left panel) shows the ratio between the terms $(2+\lambda)\int_0^{\tau}\sigma_\infty \langle\gamma_{\mathrm{map}}^2\rangle \mathrm{d}\tau'$ and $-(1+\lambda)\int_0^\tau\sigma_\infty~\mathrm{d}\tau',$ appearing in the exponent in equation (\ref{eq:norm_gamma}). It is clear that at early times the global quantity has more influence on the size of the error ${Q}_\gamma$. This is addressed in Section \ref{sec:global_err}. In contrast, at late times the global quantity is not relevant and this explains why the size of the error ${Q}_\gamma$ remains small and stable.

To evaluate the original variable $\omega(\bm{X}_\gamma(t(\tau)),t(\tau))$ from the mapped variables we use
$$ \omega(\bm{X}_\gamma(t(\tau)),t(\tau)) = \omega_{\mathrm{map}}(\bm{X}_\gamma(t(\tau)),\tau)||\gamma(\cdot,t(\tau))||_\infty. $$
We compare this against equation (\ref{om_an_map}).

\begin{figure}
\centering
\vspace{.5cm}
\hspace{5mm}
\scalebox{.9}{\input{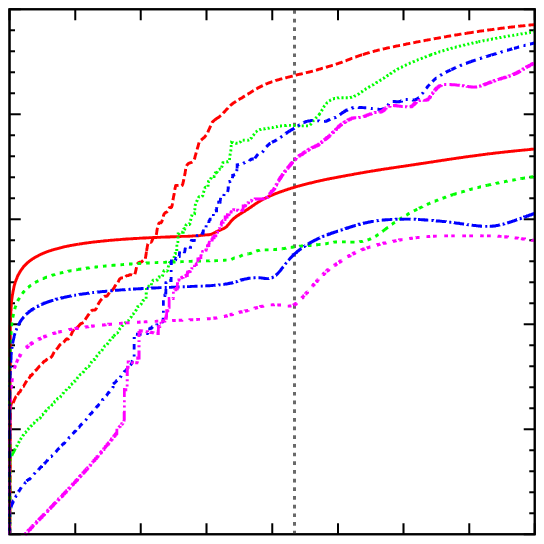}}
\hspace{1.5cm}
\vspace{.5cm}
\scalebox{.9}{\input{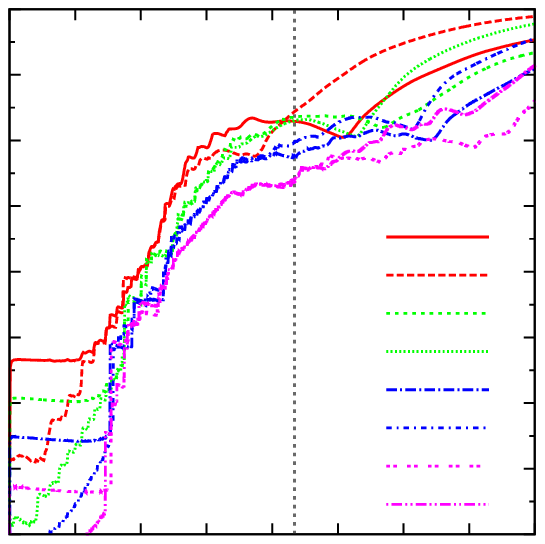}}
\hspace{-1.5cm}
\caption{Figure showing the errors $Q_\gamma$ (left) and $Q_\omega$ (right) for both the original 2D Euler model and the mapped 2D Euler model at various resolutions. Labels are (a) $N=256$ mapped system, (b) $N=256$ original system, (c) $N=512$ mapped system, (d) $N=512$ original, (e) $N=1024$ mapped, (f) $N=1024$ original, (g) $N=2048$ mapped, (h) $N=2048$ original. There is a clear accuracy gain by performing the mapping; the errors remain small for later times. Grey dotted vertical line is at $\tau=4.34$, representing the reliability time at which the $N=2048$ case becomes unresolved, see Section \ref{sec:spectra} and Table \ref{tab:Reliability}.  \label{ERROR2D5}
}
\end{figure}

Figure \ref{ERROR2D5} shows a comparison of the errors $Q_\gamma$ and $Q_\omega$ at various resolutions and demonstrates how the analysis of the mapped system along with its numerical solution serve to improve the accuracy of the blowup quantities $\|\gamma(\cdot,t)\|_{\infty}$ and $\omega(\bm{X}_\gamma(t),t)$ near the singularity time. 

\begin{figure}
\centering
\scalebox{.5}{\input{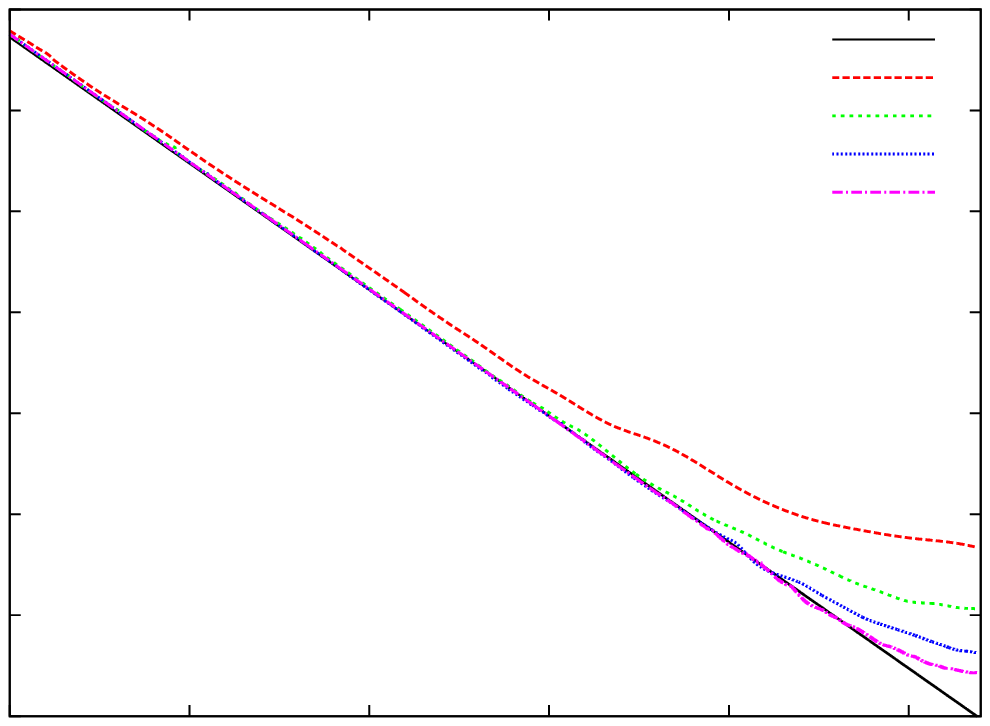}}
\scalebox{.5}{\input{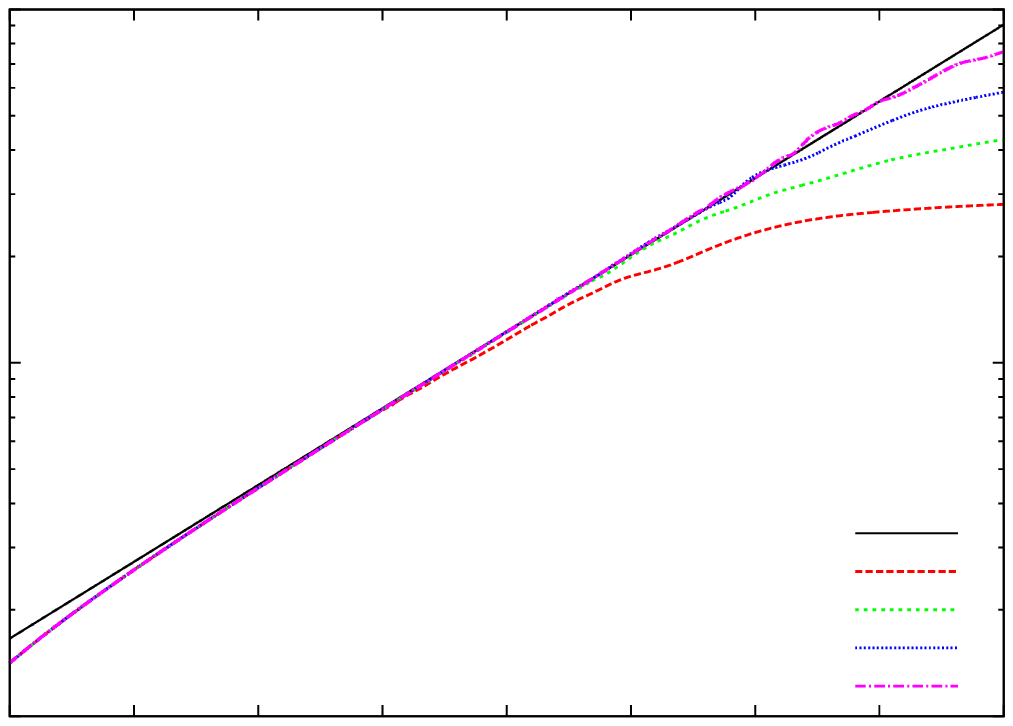}}
\caption{Two different resolution studies of the data from original system at different resolutions   $N=256, 512,1024, 2048.$ Left panel: Classical plot of $1/\|\gamma(\cdot,t)\|_\infty,$ giving a convergence to the analytically-obtained asympotic regime $1/\|\gamma(\cdot,t)\|_\infty \sim (T^*-t)/2$ (solid line). Right panel: New plot, in lin-log scaling, of $\|\gamma(\cdot,t(\tau))\|_\infty$ as a function of $\tau,$ giving a convergence to the analytically-obtained asymptotic regime $\|\gamma(\cdot,t(\tau))\|_\infty \sim \sqrt{11/2}\,\exp(\tau/2)$ (solid line).   \label{fig:ginf_orig}}
\end{figure}

We produce standard resolution convergence studies of the quantities relevant to this Section. Figure \ref{fig:ginf_orig}, left panel, shows the classical convergence study of the multiplicative inverse of the supremum norm of stretching rate, using the numerical solution of the original system, plotted as a function of original time $t$. Good convergence to the analytical result is obtained. This is the basis for the method of computing running estimates of singularity time $T^*$ (Method A in Section \ref{sec:efficient}). Figure \ref{fig:ginf_orig}, right panel, shows the less classical (but similar in spirit) lin-log convergence study of the supremum norm of stretching rate, again using the numerical solution of the original system, plotted as a function of mapped time $\tau.$ Again, good convergence to the analytical asymptote is obtained. Finally, figure \ref{fig:ratio} (right panel) shows the lin-log convergence study of the spatial average of the square of the mapped stretching rate, $\langle \gamma_{\text{map}}^2 \rangle,$ using the numerical solution of the mapped system. Good convergence to the analytical asymptote is verified.

\subsubsection{Errors in global quantities near blowup: $\langle\gamma_{\mathrm{map}}^2\rangle$ and $\langle\gamma^2\rangle$}
\label{sec:global_err}
Figure \ref{ERROR2D5} shows that the numerical solution of the mapped system contains higher early-time errors in $Q_\gamma$ than the numerical solution of the original system. These errors do not affect the late-time (near singularity) behaviour and can be controlled by reducing the time step $d\tau.$ They arise because the mapped equations contain additional terms (those proportional to $\sigma_\infty$ in equations (\ref{PDE1})--(\ref{PDE2})) which introduce an extra time scale in the $\tau$ variable, proportional to $\langle\gamma_{\mathrm{map}}^2\rangle^{-1}.$ This time scale is bounded from below and goes to infinity as $\tau\to \infty$, so we are able to resolve it by reducing the time step $d\tau$ at early times. This extra time scale feeds into the error $Q_\gamma$ via formula (\ref{eq:norm_gamma}) which gives the supremum norm of stretching rate in terms of the mapped variables. This entails the numerical approximation of the integral $\int_0^\tau \langle \gamma_{\mathrm{map}}^2\rangle \mathrm{d}\tau'$ which is sensitive to the extra time scale at early times. Figure \ref{fig:ratio} (left panel) gives a quantitative measure of the significance of this integral term as a function of time. 

We conclude that after $\tau \approx 6$ the integral term contributes less than $5\%$ to the total exponent in equation (\ref{eq:norm_gamma}). Therefore, at late times, the assessment of $\|\gamma\|_{\infty}$ using the mapped system's numerical solution is controlled by the term $-(1+\lambda)\int_0^\tau\sigma_\infty~\mathrm{d}\tau'$ in equation (\ref{eq:norm_gamma}), which surprisingly does not depend on the numerical field $\gamma_{\text{map}},$ except through the term $\sigma_\infty$ which takes values $\pm1$ according to its definition in equation (\ref{eq:sigma}). Recall that, analytically, in the case $\lambda=-3/2$ we have $\sigma_\infty = 1$ for all times, for the choice of initial conditions that we made. Hence the dominant term in the exponent is just $-(1+\lambda) \tau$ which, though still numerical, is a prescribed function of the timesteps. The role of $\sigma_\infty$ computed numerically is illustrated by looking at the error $Q_\gamma,$ figure \ref{ERROR2D5} first panel. For example, at resolution $2048,$ the jump observed at $\tau \approx 16$ is due to the fact that the numerical solution becomes under-resolved already at $\tau \approx 6,$ and consequently the quantity $\sigma_\infty(\tau)$ becomes noisy. We will discuss in detail the loss of spectral resolution of the numerical solutions in Section \ref{sec:spectra}.\\

We now perform a direct analysis of the errors in the global quantities $\langle\gamma_{\mathrm{map}}(\cdot,\tau)^2\rangle$ and $\langle\gamma^2(\cdot,t)\rangle$, computed respectively from the numerical integrations of the mapped and original systems. We compute the error in the function
$$\langle\gamma_{\mathrm{map}}(\cdot,\tau)^2\rangle = \frac{\langle\gamma(\cdot,t(\tau))^2\rangle}{\|\gamma(\cdot,t(\tau))\|_{\infty}^2}$$
 in two different ways:
 \begin{itemize}
 \item Using the numerical solution of the original system: 
 
\begin{equation}
\label{eq:Q*}
 Q_{\langle\gamma_{\mathrm{map}}^2\rangle}^* = Q\bigg(\frac{\langle\gamma_{\mathrm{num}}(\cdot,t(\tau))^2\rangle}{\|\gamma_{\mathrm{num}}(\cdot,t(\tau))\|_{\infty}^2},\frac{\langle\gamma_{\mathrm{ana}}(\cdot,t(\tau))^2\rangle}{\|\gamma_{\mathrm{ana}}(\cdot,t(\tau))\|_{\infty}^2}\bigg),
 \end{equation}

\item Using the numerical solution of the mapped system: 
 \begin{equation}
\label{eq:Qmap}
Q_{\langle\gamma_{\mathrm{map}}^2\rangle} = Q\bigg(\langle\gamma_{\mathrm{map,num}}(\cdot,\tau)^2\rangle,\frac{\langle\gamma_{\mathrm{ana}}(\cdot,t(\tau))^2\rangle}{\|\gamma_{\mathrm{ana}}(\cdot,t(\tau))\|_{\infty}^2}\bigg).
\end{equation}
 \end{itemize}
 
\begin{figure}
\centering
\input{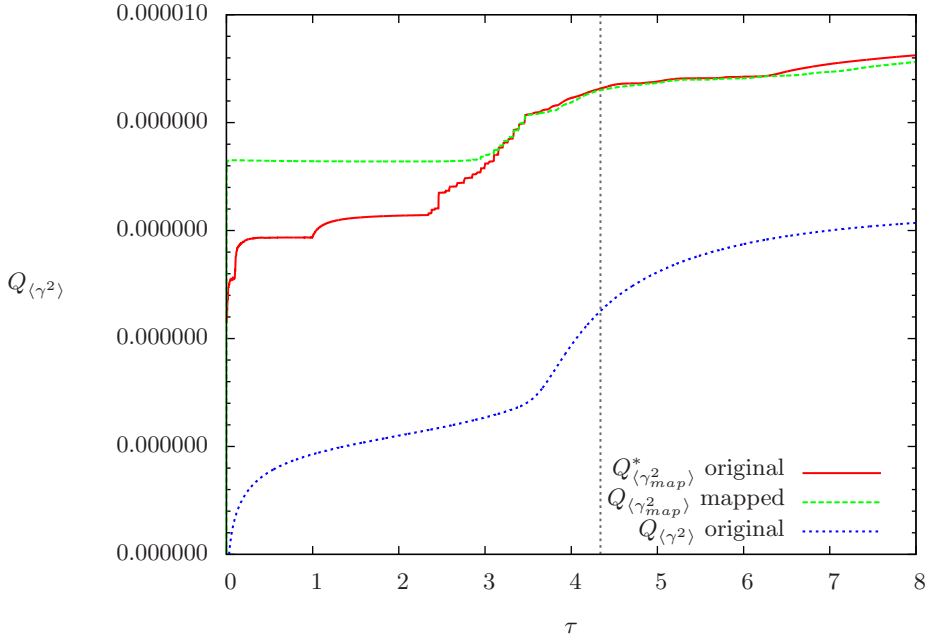}
\caption{Evolution of errors of global quantities $\langle\gamma^2\rangle$ and $\langle\gamma_{\text{map}}^2\rangle$ at resolution $N=1024$, as a function of the total integration time $\tau$, obtained from equations (\ref{eq:Q*}), (\ref{eq:Qmap}) and (\ref{eq:Qorig}). Notice that the curves $Q_{\langle\gamma_{\mathrm{map}}^2\rangle}^*$ (obtained from original system numerical data) and $Q_{\langle\gamma_{\mathrm{map}}^2\rangle}$ (obtained from mapped system numerical data) are nearly the same, illustrating that their source of errors is the same. Grey dotted vertical line is at $\tau=4.34$, representing the reliability time at which the $N=2048$ case becomes unresolved, see Section \ref{sec:spectra} and Table \ref{tab:Reliability}\label{fig:error_gsq}}
\end{figure}

Figure \ref{fig:error_gsq} shows the evolution of these errors. It is evident that they have a comparable size and behaviour. To understand this we also plot in the same figure the error
\begin{equation}
\label{eq:Qorig}
Q_{\langle\gamma^2\rangle} = Q(\langle\gamma_{\mathrm{num}}^2(\cdot,t(\tau))\rangle,\langle\gamma_{\mathrm{ana}}(\cdot,t(\tau))^2\rangle),
\end{equation}
computed directly from the numerical solution of the original system. This latter error is surprisingly small at all times, which implies that the error $Q_{\langle\gamma_{\mathrm{map}}^2\rangle}^*$ is dominated by the error in $\|\gamma_{\mathrm{num}}(\cdot,t(\tau))\|_{\infty}^2.$ On the other hand, the error $Q_{\langle\gamma_{\mathrm{map}}^2\rangle}$ includes accumulated errors due to repeated normalisations of the field $\gamma_{\text{map}}$ in the mapped system. The fact that the two errors $Q_{\langle\gamma_{\mathrm{map}}^2\rangle}$ and $Q_{\langle\gamma_{\mathrm{map}}^2\rangle}^*$ are comparable indicates that they have the same origin: $\|\gamma_{\mathrm{num}}(\cdot,t(\tau))\|_{\infty}.$

A last comment about the global quantity $\langle\gamma(\cdot,t)^2\rangle.$ In the case $\lambda = -3/2$ an interesting coincidence occurs whereby this quantity is a constant of the motion. This can be verified directly from the evolution equation (\ref{PDE1}). One may think of this situation as resembling 3D Euler, where energy is conserved numerically even at late times when resolution has been lost. The fact that our error $Q_{\langle\gamma^2\rangle}$ remains consistently small for all times is thus not a surprise, rather a consequence of the pseudo-spectral method (this is highlighted in figure \ref{fig:error_gsq} by the vertical line denoting the reliability time defined in section \ref{sec:spectra}). For other values of $\lambda,$ we expect (and this will be demonstrated in a forthcoming paper) that the error $Q_{\langle\gamma^2\rangle}$ grows without bound due to loss of resolution.

One should be careful about drawing too strong conclusions from these error measures. It should be remembered that the mapping does nothing to improve the spatial resolution of a calculation, so any small scale structures present in the flow may be expected to suffer from a loss of resolution at roughly the same time. How this contributes to the errors in certain measures of the flow will vary from one case to the next and from one measure to the next. To investigate this we consider the spectra of $\gamma$ in the following section.

\subsection{Spectra, analyticity strip and Beale-Kato-Majda (BKM) theorem}
\label{sec:spectra}

An effective diagnostic for the spatial collapse associated with a typical finite-time blow-up scenario presents itself in the form of the analyticity strip \citep{Sulem:1983,bustamante2012interplay}. Given the spectrum of spatial Fourier coefficients provided in our numerical simulation, the $L^2$ spectrum of  $\gamma$ is defined as the sum of the squares of modulus of the Fourier coefficients over circular shells, or in short, the $L^2$ stretching-rate spectrum:
$$E(k,t)=\sum_{k-\frac{1}{2}<\left|\textbf{k}\right|< k+\frac{1}{2}}|\hat{\gamma}(\textbf{k},t)|^2.$$
While we know analytical solutions for the blowup quantities, so far we have not found a method to obtain analytical expressions regarding the stretching-rate spectrum, other than the formula valid for $\lambda=-3/2,$ stating $\sum_{k=1}^\infty E(k,t) = 3/4 \,\,\text{(const)}.$ 

The lack of analytical results for spectra is seen as an advantage since it allows us to test the analyticity-strip method and its bridge with the BKM theorem \citep{bustamante2012interplay} from a purely numerical point of view. The results presented in this section can therefore be contrasted against future analytical developments.

\begin{figure}
\centering
\scalebox{.5}{\input{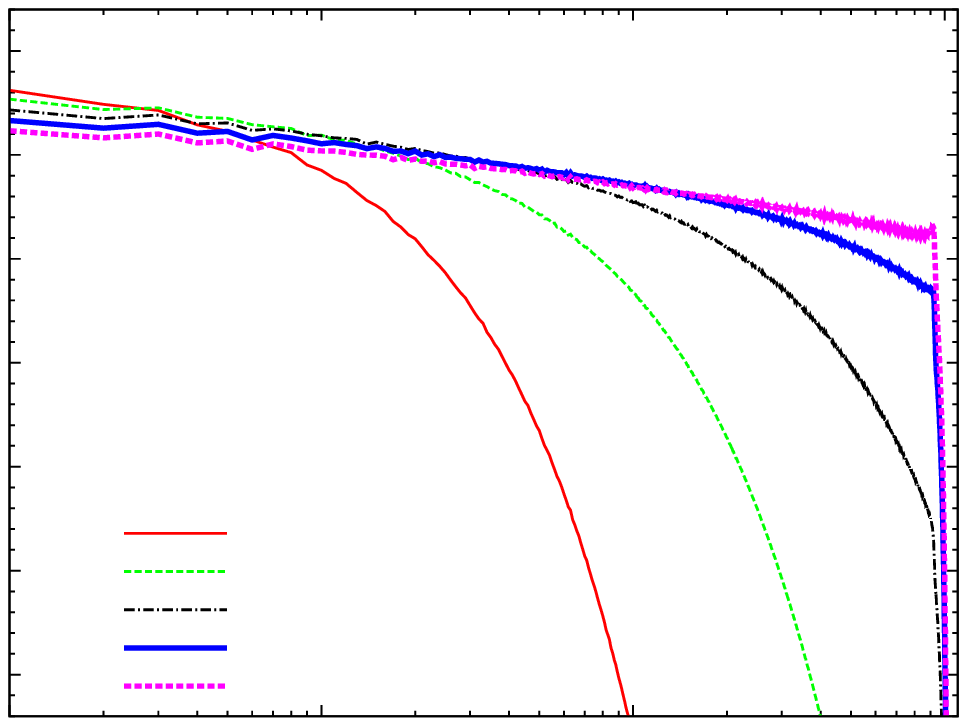}}
\scalebox{.5}{\input{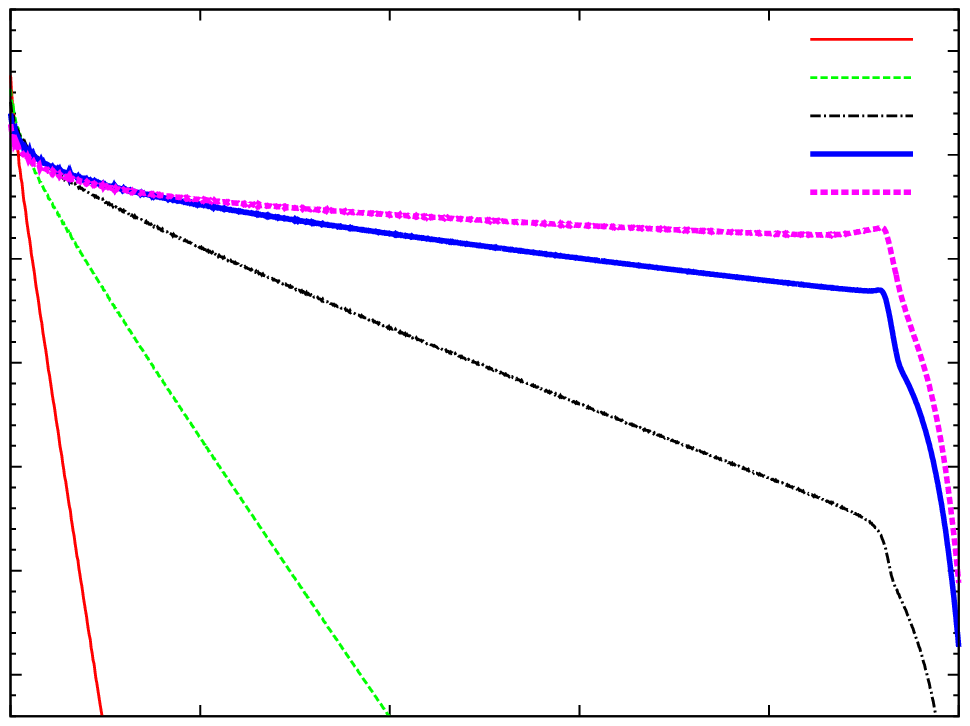}}
\caption{Left and right panels: Snapshots of stretching-rate 1D shell spectra $E(k,t)$ at mapped times $\tau=1,2,3,4,5$ (curves progressing from bottom to top) in log-log scale (left panel) and lin-log scale (right panel).  The slopes of the straight lines on the left panel are proportional to the exponent $n_E(t)$. The slopes of the straight lines on the right panel are proportional to the logarithmic decrement $\delta_E(t).$ Resolution: $N=2048.$  \label{fig:1Dspectra}}
\end{figure}

The function $\gamma(\mathbf{x},t)$ remains analytic in the space variables if $E(k,t)$ can be bounded by
$$E(k,t) \lessapprox C_E(t) k^{-n_E(t)} \text{e}^{-2k\delta_E(t)}\,,$$
where $\delta_E(t)$ is the analyticity strip width, also known as the logarithmic decrement, and $C_E(t), n_E(t)$ are positive numbers. We assume this approximation holds for our functions. 
Figure \ref{fig:1Dspectra} shows snapshots of the $L^2$ spectrum $E(k,t)$, in log-log as well as lin-log scaling, to provide evidence of the feasibility of this approximation. The common procedure is to find the coefficients $C_E(t), n_E(t), \delta_E(t)$ by performing a least-squares fit, at each time $t,$ on $\ln E(k,t)$ over some interval $k_i \leq k \leq k_f.$ The problem becomes linear in the parameters $\ln C_E(t), n_E(t)$ and $\delta_E(t).$ More details can be found in, e.g., \citep{bustamante2012interplay}, equation 5. 

\begin{figure}
\centering
\scalebox{.5}{\input{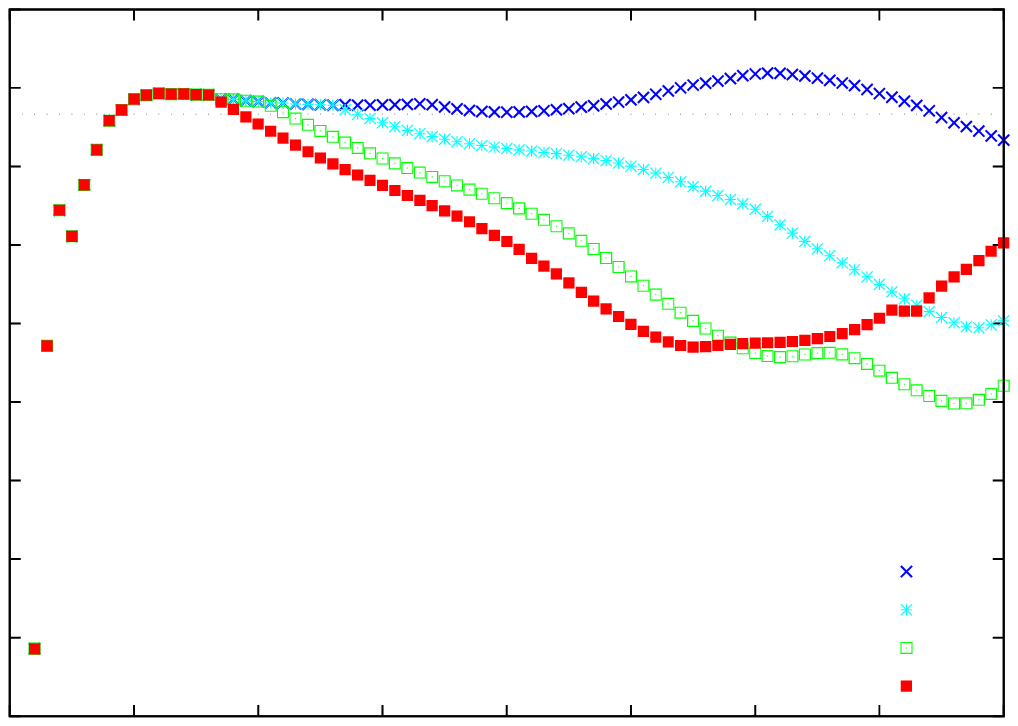}}
\hspace{-5mm}
\scalebox{.5}{\input{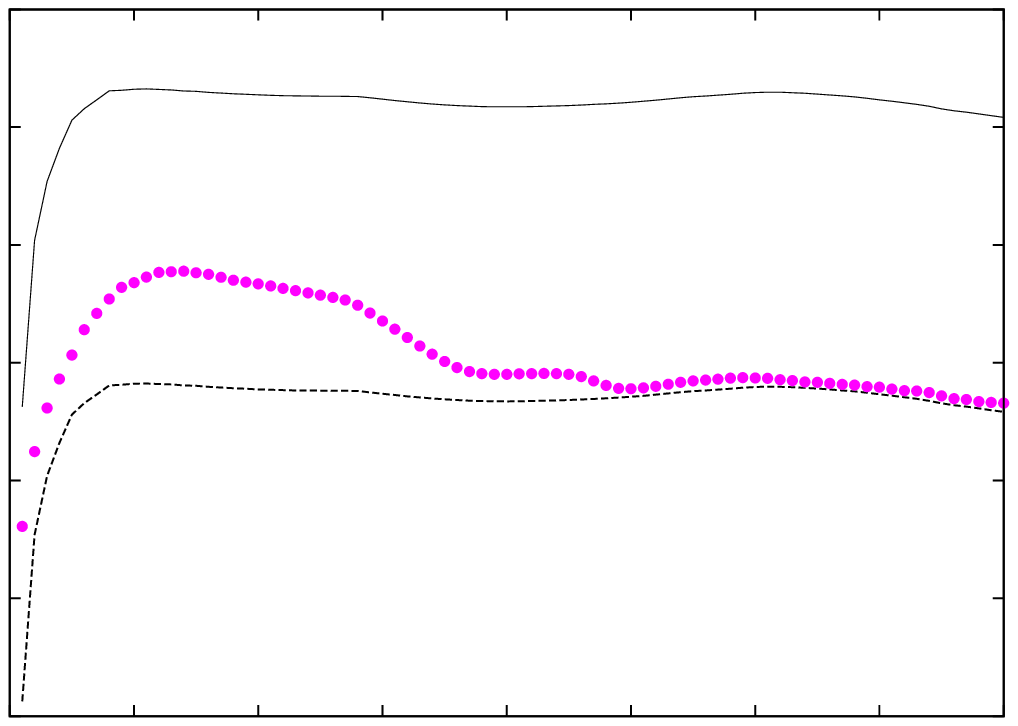}}
\hspace{-5mm}
\scalebox{.5}{\input{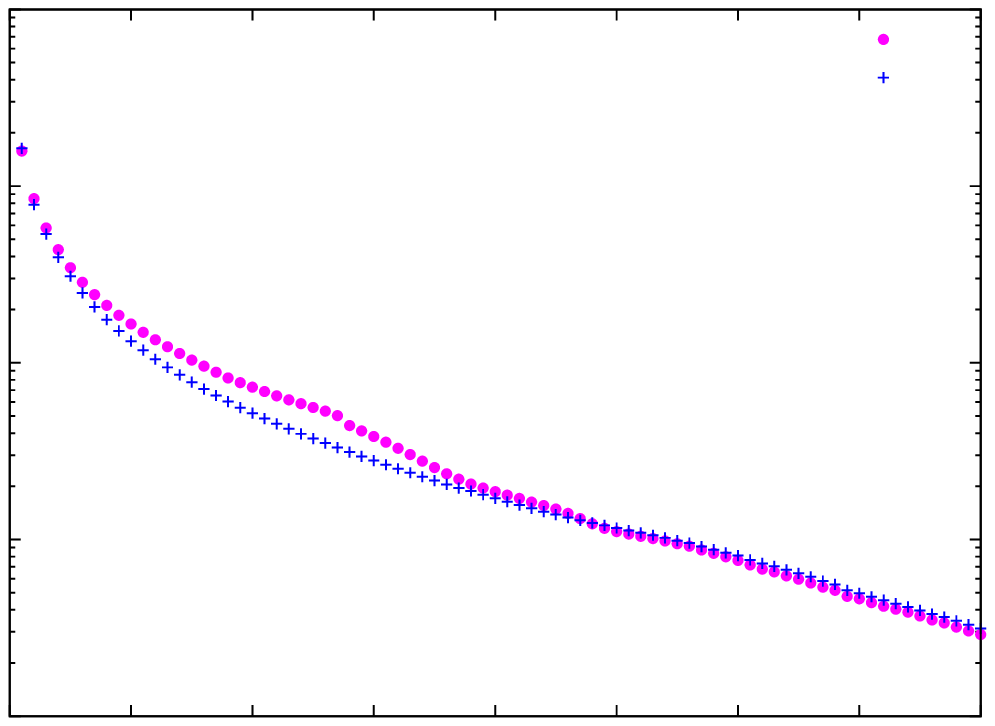}}
\hspace{-5mm}
\scalebox{.5}{\input{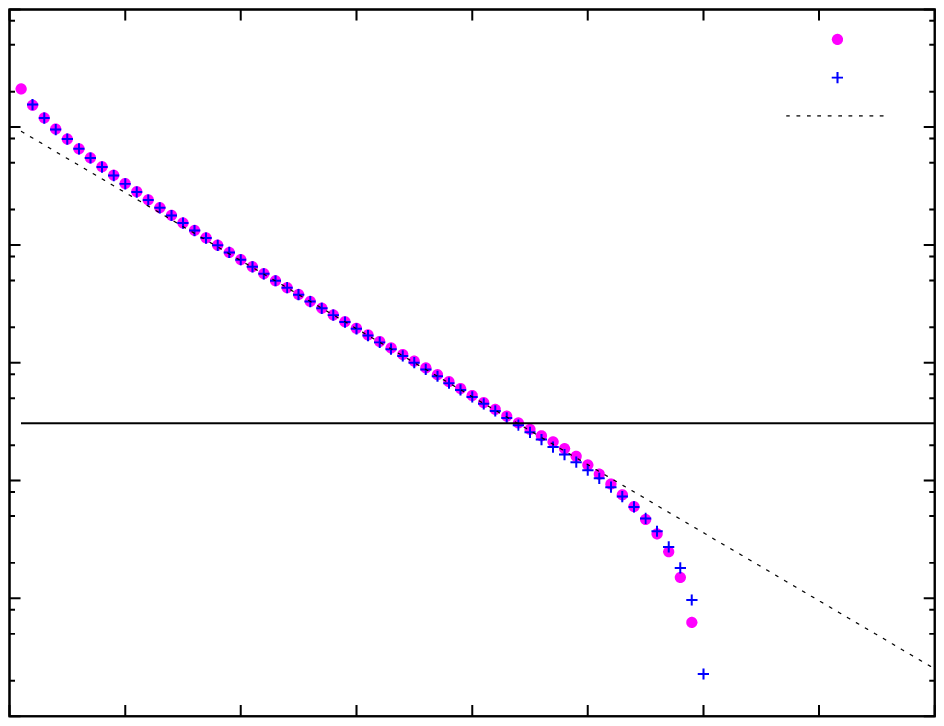}}
\caption{Results at resolution $N=2048$ (except top left panel) for the fit parameters of the $L^2$ spectra $E(k,t)$ and $L^1$ spectra $F(k,t)$ at several times (as function of mapped time $\tau$). Top left panel: Resolution study ($N=256,512,1024,2048$) for $n_E(t(\tau)).$ Progressive convergence is observed towards $n_E \approx 5/3$ (within a $5\%$ error) near reliability time.  Top right panel: Results for $n_F(t(\tau))$ (circles). Solid and dashed curves represent the upper and lower bounds $n_E/2$ and $(n_E - 1)/2,$ respectively, cf. inequality (\ref{eq:sandwich}). The curve $n_F$ is consistently between these bounds. Bottom left panel: Results for $C_E(t(\tau))$ ($+$ symbols) and $C_F(t(\tau))$ (circles). At late times these coincide, thus confirming the isotropy of the 2D spectrum. Bottom right panel: Results for $\delta_E(t(\tau))$  ($+$ symbols) and $\delta_F(t(\tau))$ (circles). These coincide, in agreement with inequalities (\ref{eq:sandwich}). The horizontal line is the smallest resolved scale $\Delta x = 2\pi / 2048$ and the dotted line corresponds to the numerical fit $\delta = \exp (0.53 - 1.33 \tau)$ obtained using data in the range $3\leq \tau\leq 4.34.$  \label{fig:n_and_delta_tau_2}}
\end{figure}

It is customary to define a ``reliability time barrier'' $t_{\mathrm{rel}}$ by the condition 
$$\delta_E(t) \leq dx \Leftrightarrow t \leq t_{\mathrm{rel}}\,,$$
 where $dx$ is the grid spacing of the numerical simulation. This barrier represents the obvious requirement that the smallest scales available in the numerical simulation are well resolved.  
 Figure \ref{fig:n_and_delta_tau_2} shows the results at resolution $N=2048$ for the fit parameters at several times (in mapped time $\tau$). Right panel indicates a reliability time $\tau_{\text{rel}} \approx 4.2$, corresponding to $t_{\text{rel}} \approx 1.121.$  Left panel, $+$ symbol, shows $n_E(t)$ with the remarkable convergence to $n_E = 5/3 \pm 0.08$ (dotted horizontal line) near reliability time. The error bar $0.08$ stands for the $5\%$ error of the least-squares fit procedure.

A classical method used in \citep{bustamante2012interplay} gives rise to the following rigorous inequality:
$$\|\gamma(\cdot,t)\|_\infty \leq \sum_{k=1}^\infty \sum_{k-\frac{1}{2}<\left|\textbf{k}\right|< k+\frac{1}{2}} |\hat{\gamma}(\textbf{k},t)|.$$
This inequality is saturated if and only if there is alignment of the phases of the Fourier components $\hat{\gamma}(\textbf{k},t)$ that carry a significant amplitude. This alignment is expected to happen near the singularity time, as learned from the 1D inviscid Burgers equation. The fact that a $L^1$ norm (rather than the more familiar $L^2$ norm $E(k,t)$) appears in this rigorous inequality, motivates the introduction of a new type of spectrum, the $L^1$ stretching-rate spectrum
$$F(k,t) = \sum_{k-\frac{1}{2}<\left|\textbf{k}\right|< k+\frac{1}{2}} |\hat{\gamma}(\textbf{k},t)|\,.$$
In terms of this $L^1$ spectrum the above inequality becomes 
\begin{equation}
\label{eq:good_ineq}
\|\gamma(\cdot,t)\|_\infty \leq \sum_{k=1}^\infty F(k,t).
\end{equation}

We now make a connection between this new $L^1$ spectrum and the more familiar $L^2$ spectrum, via a rigorous equivalence:
\begin{equation}
\label{eq:sandwich}
\sqrt{E(k,t)} \leq F(k,t) \leq \sqrt{S_k} \sqrt{E(k,t)}\,,
\end{equation}
where 
$$S_k \equiv \sum_{k-\frac{1}{2}<\left|\textbf{k}\right|< k+\frac{1}{2}} 1.$$
We have the approximate result valid as $k\to\infty$:
 $$S_k \approx 2 \pi k\,.$$
Inequalities (\ref{eq:sandwich}) allow us to work with the $L^1$ spectrum $F(k,t)$ in the same way as we would work with the more familiar $L^2$ spectrum $E(k,t).$ In particular, a fit of the form
$$E(k,t) \lessapprox C_E(t) k^{-n_E(t)} \exp(-2 k \delta_E(t))$$
implies a fit of the form
$$F(k,t) \lessapprox C_F(t) k^{-n_F(t)} \exp(- k \delta_F(t)).$$
One could interpret these two fits as working hypotheses, as in \citep{bustamante2012interplay}. In the limit $k\to \infty$ the inequalities (\ref{eq:sandwich}) imply a sandwich so $\delta_E(t) = \delta_F(t)$ is necessary. This is verified numerically in Fig.\ref{fig:n_and_delta_tau_2}, bottom right panel.

In the intermediate-$k$ range, the inequalities imply the following bounds:

\begin{equation}
\label{eq:bounds_n_F}
\frac{n_E(t) -1}{2} \leq n_F(t) \leq \frac{n_E(t)}{2}\,.
\end{equation}
These bounds are verified in Fig.\ref{fig:n_and_delta_tau_2}, top right panel.

\begin{figure}
\centering
\scalebox{.8}{\input{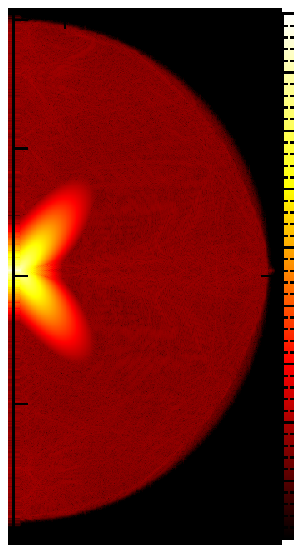}}
\hspace{2.2cm}
\scalebox{.8}{\input{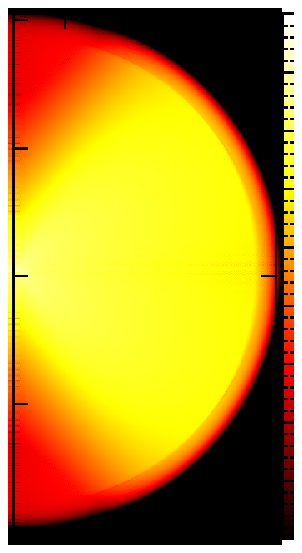}}
\caption{Snapshots of stretching-rate 2D spectra at mapped times $\tau=2$ (left panel) and $\tau=4$ (right panel). Resolution: $N=2048.$  \label{fig:2Dspectra}}
\end{figure}

Finally, the proportionality factors $C_F(t)$ and $C_E(t)$ cannot be easily related unless the above bounds (\ref{eq:bounds_n_F}) for $n_F(t)$ become saturated. For example, if the upper bound was saturated, this would correspond rigorously to a very anisotropic $2D$ spectrum $|\hat{\gamma}(\textbf{k},t)|$ in terms of orientation of $\mbf{k}.$ In this case we would have  $C_F \approx \sqrt{C_E}.$
On the other hand, if  the lower bound was saturated, as it happens at late times (see Fig.\ref{fig:n_and_delta_tau_2}, top right panel) this would correspond rigorously to a very isotropic $2D$ spectrum $|\hat{\gamma}(\textbf{k},t)|.$ In this case we would have  $C_F \approx \sqrt{2 \pi C_E}.$ Figure \ref{fig:n_and_delta_tau_2}, bottom left panel, shows the curves $C_F$ and $\sqrt{2 \pi C_E}$ as functions of time, showing equality at late times. So we can conclude that the $2D$ spectrum becomes nearly isotropic at late times.  To support this analysis we provide in figure \ref{fig:2Dspectra} two snapshots of the stretching-rate 2D spectrum, at mapped times $\tau = 2$ (left panel) and $\tau=4$ (right panel). It is evident that the spectrum is strongly anisotropic at early times and evolves towards isotropy at late times, in agreement with the saturation of inequalities found.

Following an analogous discussion to that in \citep{bustamante2012interplay}, we see that the left-hand-side of inequality (\ref{eq:good_ineq}) has a singular behaviour. In fact, a BKM-type of theorem can be demonstrated for the left-hand side, namely we can assume
$$\int_0^{T^*} \|\gamma(\cdot,t)\|_\infty dt = \infty.$$
This is obvious from the analytical solution presented in Table \ref{Cases}. This  will imply a singular behaviour of the right-hand side of inequality (\ref{eq:good_ineq}):
$$\int_0^{T^*} \sum_{k=1}^{\infty}F(k,t)  dt = \infty.$$
Using the above fit for $F(k,t)$ we get the result
$$\int_0^{T^*} \sum_{k=1}^{\infty}k^{-n_F(t)} \exp(- k \delta_F(t))  dt = \infty.$$
We recognise the Jonquiere's function $\text{Li}({n_F(t)},\text{e}^{-\delta_F(t)}).$ We get
$$\int_0^{T^*} \text{Li}({n_F(t)},\text{e}^{-\delta_F(t)}) dt= \infty.$$
Now, in the limit as $t \to T^*$ we have $\delta_F \to 0$ so we can approximate the Jonquiere's function to get
$$\int_0^{T^*} (\delta_F(t))^{n_0-1} dt= \infty,$$
where $n_0 = \liminf_{t \to T^*} n_F(t).$ So the asymptotic behaviour $\delta_F(t) \sim (T^*-t)^\Gamma$ would be consistent with singularity behaviour if and only if
$$\Gamma \geq \frac{1}{1-n_0}.$$

From Fig.\ref{fig:n_and_delta_tau_2}, left panel, we see that $n_0 \approx 0.36$ if we consider the data near reliability time ($\tau_{\text{rel}}\approx 4.34$). This gives $\Gamma \geq 1.56,$
which is consistent with the fits obtained from Fig.\ref{fig:n_and_delta_tau_2}, right panel, that produce  $\Gamma \approx 2.66$ by virtue of the analytical result $(T^*-t) \approx \exp(-\tau/2).$ At the same time this shows that the inequality (\ref{eq:good_ineq}) is not saturated by the field $\gamma(\mbf{x},t).$ The interpretation of this lack of saturation is that the Fourier phases, 
$\{\arg \hat{\gamma}(\textbf{k},t)\}_{k=1}^\infty,$ do not all align near the singularity time. In fact this is evident from the physical-space snapshots in Fig. \ref{fig:2Dplots}, where the singular structure is between a filament and a point.

\subsection{Estimating the singularity time $T^*$ efficiently}
\label{sec:efficient}

It would be meaningless to provide results comparing the accuracy of particular numerical methods, without including some quantification of their relative computational expense. 

It is possible to obtain two independent estimates of the singularity time $T^*$ by using the numerical solutions from either the original system or the mapped system:

\begin{enumerate}
\item[\textbf{Method A.}]
For the original system, we fit the stretching rate norm $\|\gamma(\cdot,t)\|_{\infty}$ as a power law locally in time, using a method introduced in \citep{Bustamante:2008p2289} to determine an estimate of singularity time $T^*$. The power-law fits are of the form
\begin{equation}
f(t) \propto (T_{\text{A}}^*-t)^{\alpha},
\label{FIT1}
\end{equation}
where $f(t)$ stands for $\|\gamma(\cdot,t)\|_{\infty}$ in this case. This ansatz is justified in this case by the analytically obtainable asymptotic formulae in Table \ref{Cases}. The local fits are achieved using the function
\begin{equation}
g(t)=\left(\frac{d \ln f(t) }{dt}\right)^{-1} =\frac{ f }{\dot{f}} =- \frac{1 }{\alpha} (T_{\text{est}}^*-t).
\label{FIT2}
\end{equation}

Instantaneous running estimates $\alpha$ and $T_{\text{A}}^*$ are then computed by linear-fitting the function $g(t)$ instantaneously using adjacent data points, or more generally over a small time window  (of size $\sim 0.2$) containing a good number of data points, in order to eliminate spurious oscillations in the running estimates. Notice that this method provides an extra quantity: the exponent $\alpha$, which serves as an extra measure of validation. In the case $\lambda = -3/2$ one should get $\alpha = -1$ (see Table \ref{Cases}). This validation is consistently held throughout the computation (figure not shown).

\item[\textbf{Method B.}]
For the mapped system, the situation relies on the explicit formula (\ref{eq:T_star}). There, the only relevant numerical quantity is $\int_0^\tau\langle\gamma_{\mathrm{map}}(\cdot, \tau')^2\rangle d\tau'.$ In analogy to the previous case, we will fit the integrand. But, unlike the previous case, we cannot rely on local fits because doing this would lead to accumulation of errors in the estimation of the integral. 

Using equations (\ref{eq:norm_gamma})--(\ref{eq:T_star}) we obtain the estimate
\begin{eqnarray}
\nonumber
T_{\text{B}}^*(\tau) &=& \int_0^\infty \frac{1}{\|\gamma(\cdot,t(\tau'))\|_\infty}d\tau'\\
 &\approx& \left(\int_0^\tau \frac{1}{\|\gamma(\cdot,t(\tau'))\|_\infty}d\tau'\right)_{\text{num}} + \left(\int_\tau^{\widehat{\tau}} \frac{1}{\|\gamma(\cdot,t(\tau'))\|_\infty}d\tau'\right)_{\text{extrap}}
\end{eqnarray}
where in both terms we compute the integrand using equation (\ref{eq:norm_gamma}). The subscript ``num'' means that we use the numerical solution of the mapped system to compute the time integrals, using Simpson's rule. As for the subscript ``extrap'', $\widehat{\tau}$ is a very big number chosen so that the numerical integral converges  ($\sim 1000$ in practice) and we compute the integrand, using (\ref{eq:norm_gamma}), as follows:
$$\frac{1}{\|\gamma(\cdot,t(\tau'))\|_\infty} = \frac{1}{\|\gamma(\cdot,t(\tau))\|_\infty} 
\exp\left[(1+\lambda)\int_\tau^{\tau'}\sigma_\infty~\mathrm{d}\tau'' - (2+\lambda)\int_\tau^{\tau'}\sigma_\infty \langle\gamma_{\mathrm{map}}^2\rangle \mathrm{d}\tau''\right]$$
where we set $\sigma_{\infty}(\tau'') = \sigma_{\infty}({\tau})$ in the above exponent, and we model the functions appearing in the above exponent using the following fit ansatz that is motivated by the generic asymptotic behaviour of $\langle\gamma_{\mathrm{map}}^2\rangle$ as $\tau'' \to \infty:$
$$\langle\gamma_{\mathrm{map}}^2(\cdot,\tau'')\rangle  = \frac{\beta}{2+\lambda}\left[c \exp(\beta \tau'')-1\right]^{-1}\,,$$
where $\beta$ and $c$ are two positive fit parameters. In order to obtain these fit parameters we use the numerical data for $\langle\gamma_{\mathrm{map}}^2(\cdot,\tau'')\rangle$ in the range $0 \leq \tau'' \leq \tau$ and use a least-squares fit. The integral $\int_\tau^{\tau'}\sigma_\infty \langle\gamma_{\mathrm{map}}^2\rangle \mathrm{d}\tau''$ is done analytically in terms of the fit parameters. Finally, the resulting integral $\int_\tau^{\widehat{\tau}} \frac{1}{\|\gamma(\cdot,t(\tau'))\|_\infty}d\tau'$ is done  using Simpson's rule.

\end{enumerate}

It is important to stress that in both original and mapped systems the estimation of the singularity time $T^*$ depends on two fit parameters. In general, we tried to lever as much accuracy as possible from both methods. For example we used adaptive time stepping in the original system to get a distribution of data points that is comparable to that of the mapped system, so that a more accurate estimate for $T^*$ could be obtained in the original system.

\begin{figure}
\centering
\input{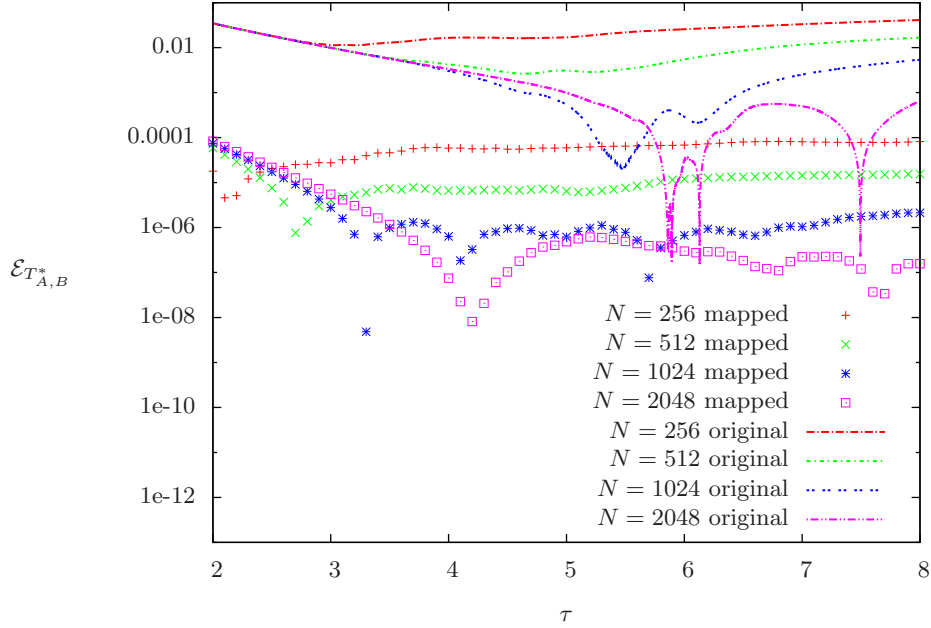}
\caption{Errors in the running estimates of singularity time $T^*$ using data from original system's numerical integration (lines) and mapped system's numerical integration (symbols) at different resolutions: from top to bottom in each case, $N=256, 512,1024, 2048.$  \label{fig:Tstar_running}}
\end{figure}

We present results from original and mapped systems regarding the assessment of estimates of singularity times. First, in figure \ref{fig:Tstar_running} we plot the relative error of the running estimates 
$${\mathcal{E}}_{T_{\text{A,B}}^*}(\tau) = \left|\frac{T_{\text{A,B}}^*}{T_{\text{ana}}^*} -1 \right|$$
where $T_{\text{ana}}^* = \frac{4}{\sqrt{3}}\arctan\left(\frac{\sqrt{6}}{4}\right) \approx 1.26894$ is the analytically-computed singularity time and $T_{\text{A,B}}^*$ stands for the running estimate obtained either from Method A (for the original system) or from Method B (for the mapped system). It is observed from the figure that: (i) For each method, there is good resolution convergence in the assessment of $T^*.$ (ii) Method B (for the mapped system) produces much better results as compared to Method A (for the original system), with an improvement of about three orders of magnitude at any given resolution.

The second set of results is the following. We produce, from each method, a single estimate (not a running estimate) $T_{\text{A,B}}^0$ of the singularity time, computed using the running estimates already obtained. It is important to stress the perhaps obvious fact that the procedure to find this single estimate is completely independent of any previous knowledge of the singularity time $T^*$. The procedure is simple: since we have a reliability time $t_{\text{rel}}$ (or, in mapped time, $\tau_{\text{rel}}$) well defined for each resolution, in terms of the analysis of spectra done in Section \ref{sec:spectra}, we  evaluate our running estimates at the reliability time. So we define, for the mapped system, $T_{\text{B}}^0 = T_{\text{B}}^*|_{\tau_{\text{rel}}}.$ As for the original system, setting the single estimate to $T_{\text{A}}^*|_{t_{\text{rel}}}$ would be possible, however we found that a better estimate is obtained by averaging the running estimates between $t_{\text{rel}}$ and $t_{\text{min}},$ where $t_{\text{min}} (> t_{\text{rel}})$ depends on the resolution and is defined by the time at which the running estimate has a global minimum.

\begin{figure}
\centering
\input{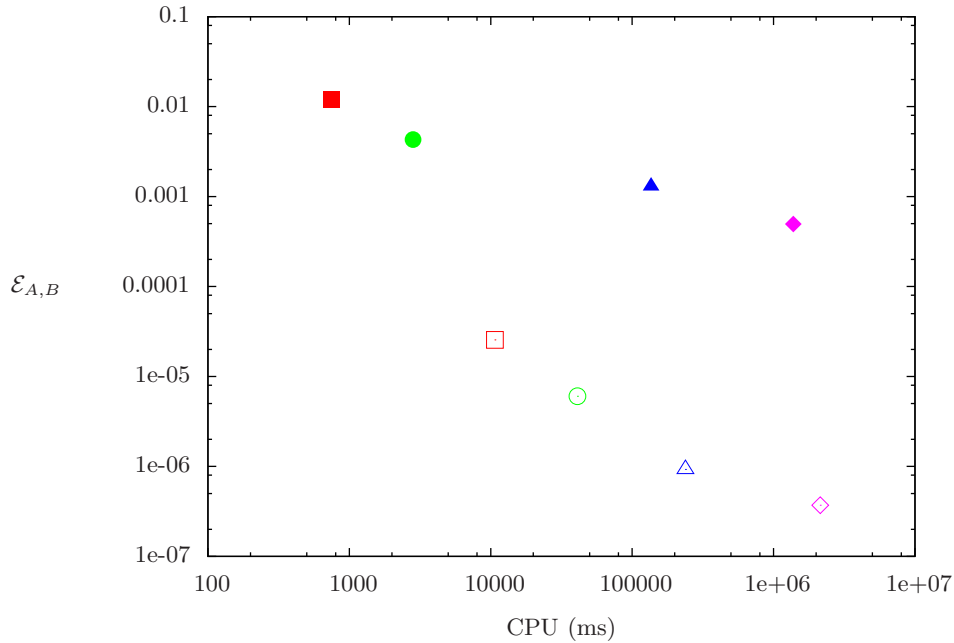}
\caption{Figure showing the singularity-time error ${\mathcal E}_{A,B}$ [error between estimated singularity time and analytically computed singularity time $T^*$, equation (\ref{eq:sing_time_error})], as a function of CPU time for both the original system, Method A (filled symbols) and the mapped system, Method B (open symbols) at various resolutions: $N=256$ (red squares), $N=512$ (green circles), $N=1024$ (blue triangles), $N=2048$ (magenta diamonds). The CPU overhead of applying the mapping is shown to be more than covered by the accuracy gain. \label{cpu_error}}
\end{figure}

Figure \ref{cpu_error} shows the CPU time versus relative error of the estimated singularity time, 
\begin{equation}
\label{eq:sing_time_error}
{\mathcal E}_{\text{A,B}} = \left|\frac{T_{\text{A,B}}^0}{T_{\text{ana}}^*}-1\right|\,,
\end{equation}
for the numerical solutions of both the original (Method A) and mapped systems (Method B) at various resolutions. It is clear that while the mapping incurs some additional expense in evaluating the extra terms, computing the interpolated supremum and applying the normalisation, it is far outweighed by the positive effect on the errors (three orders of magnitude in this study). In these measures one can make a significant improvement, saving not only CPU time, but also the memory cost of high resolution runs.

We present in Table \ref{tab:Reliability} a useful summary of the reliability times and relative errors ${\mathcal E}_{\text{A,B}}$ of the estimated singularity time, for a range of resolutions, showing that the  errors are dramatically reduced in the case of the mapped equations.

\setlength{\tabcolsep}{12pt}
\begin{table}
\centering
    \begin{tabular}{ c  c  c  c  c}
   N &  $t_{\text{rel}}$ & $\tau_{\text{rel}}$ & ${\left|T_A^0 / T^* - 1\right|}$& ${\left|T_B^0 / T^* - 1\right|}$\\ 
 \\
    $256$ & $0.9707$ & $2.8$      &  $1.2 \times 10^{-2}$ &  $2.55 \times 10^{-5}$ \vspace{2pt} \\ 
    $512$ & $1.041$ & $3.34  $    &  $4.3\times 10^{-3} $&   $6.03 \times 10^{-6} $\vspace{2pt} \\ 
    $1024$ & $1.0928$ & $3.85$  &  $1.3 \times 10^{-3}$&   $9.28\times 10^{-7} $\vspace{2pt} \\ 
    $2048$ & $1.13116$ & $4.34$ & $4.9 \times 10^{-4}$ &  $3.70 \times 10^{-7} $ \vspace{2pt}
    \end{tabular}
       \caption{Summary of results for a range of resolutions: reliability times, obtained using the stretching-rate spectra (original or mapped system give the same reliability times); relative error of the estimated singularity time $T_A^0$ stemming from original system's numerical data (Method A); relative error of the estimated singularity time $T_B^0$ stemming from mapped system's numerical data (Method B).}
    \label{tab:Reliability}
\end{table}
\setlength{\tabcolsep}{6pt}

\newpage

\section{Conclusion and Discussion}
\label{sec:concl}
In this paper we have introduced a new family of symmetry plane models of the 3D Euler equations and presented numerical and analytical solutions exhibiting finite-time blow-up. Although these models do not necessarily correspond to actual solutions of 3D Euler equations they still represent a valuable simplified and tuneable setting for the study and assessment of finite-time singularity in an idealised fluid. We make use of this example to evaluate the performance of the mapping to regular systems of \cite{bustamante20113d} in improving the diagnosis of singular behaviour. We simulate both systems using the same pseudospectral methods and find that direct determination of blowup quantities from the numerical integration of the mapped regular system produces more accurate and reliable results compared to the integration of the original system. 

We present a thorough investigation of the evolution of the Fourier spectrum of the numerical solution, however unlike the case of the supremum norms of the fields, there is no available explicit solution for the Fourier components. We validated the numerical solution by checking rigorous bounds on the Sobolev norms, using the working hypotheses introduced by \cite{bustamante2012interplay}. These hypotheses were designed in order to bridge the study of the loss of analyticity of solutions with the classical Beale-Kato-Majda type of theorems. The main results here are:

\begin{enumerate}
\item[(i)] The finite-time blowup of the supremum norm of the stretching rate implies that the Fourier spectrum's logarithmic decrement $\delta(t)$ (a measure of the loss of analyticity) must decay to zero fast enough at the singularity time. We observe a decay $\delta(t) \sim (T^*-t)^3,$ consistent with the rigorous bounds.
\item[(ii)] An ``inertial range'' of wavenumbers at which the conserved quantity $\langle\gamma^2\rangle$ is transferred to small scales is found, with a 1D spectrum (i.e., shell-integrated) of the form $E(k,t) \sim k^{-5/3}$ at times close to the singularity time (but still resolved).
\item[(iii)]  The 2D spectrum of the Fourier amplitude $|\hat{\gamma}(\mbf{k},t)|^2$ becomes isotropic at late times, in agreement with the saturation of our rigorous bounds for the $L^1$ shell spectrum in terms of the $L^2$ spectrum $E(k,t)$. Figures \ref{fig:n_and_delta_tau_2} and \ref{fig:2Dspectra} complement the above results.
\end{enumerate}

We discuss, in the interest of fairness, a technicality that arises in the mapped system. Recovering the original system's supremum norm from the mapped variables has a subtlety. At late times the dominant contribution comes from a term that depends explicitly on $\tau,$ which is therefore independent of the numerical simulations. This explains the strong and robust late-time convergence we observe in our comparisons. In a work in progress we will consider the case $\lambda \in [-1,0],$ where this behaviour is relaxed.

On the other hand, the coincidence occurring at $\lambda = -3/2$ where  $\langle \gamma^2\rangle$ is an invariant of the motion, reduces the errors in the original system with respect to the mapped system. In a work in progress we have confirmed that for any other value of $\lambda$ this invariance does not hold and as a result the original system accumulates significant additional errors. This scenario favours the mapped system even more than in the case $\lambda = -3/2$ studied here.

To estimate the singularity time we perform a two parameter fit for both the original and mapped systems in order to \emph{extrapolate} a running estimate for the singularity time. The result is up to three orders of magnitude increase in accuracy when employing the mapping over the original system. It should be emphasised that this gain in performance stems from a number of sources, for example, the form of the extrapolation to compute $T^*$, the global quantity $\langle \gamma_{\mathrm{map}}^2\rangle$ being used to ``unmap" the variables, the redistribution of numerical error within the simulation via the normalisation procedure and finally the mapping of time to distribute data appropriately near $T^*$. On this final point, it is tempting to assume that this is the main advantage of the mapping and it would be equivalent to simply employ an adaptive time step. This paper has shown, not only are the accuracy gains unrelated to time step convergence, but also that the manner of recomputing the original variables has important consequences. For these reasons, and the observations summarised earlier in this Section, we show errors (figures \ref{ERROR2D5} and \ref{fig:error_gsq}), singularity time estimates $T^*$ (figure \ref{fig:Tstar_running}), etc. at values of $\tau$ far beyond the usual reliability time cut-off.

To conclude this Section we remark that the work done here has served to benchmark and validate the methods that we will apply in a forthcoming paper for the analysis of numerical simulations of 3D Euler equations, in both original variables and mapped variables. In the 3D case, we do not have at hand any analytical solution to compare with. However, the fact that the mapped system's numerical solution leads to a more accurate estimation of singularity time than the original system, motivates the use of the mapped approach on the 3D Euler equations.

\section{Acknowledgments}
This publication has emanated from research supported in part under the Programme for Research in Third Level Institutions (PRTLI) Cycle 5; the European Regional Development Fund; and a research grant from Science Foundation Ireland (SFI) under Grant Number 12/IP/1491. Computational resources and support were provided by the DJEI/DES/SFI/HEA Irish Centre for High End Computing (ICHEC) under class C project  \emph{ndmat023c}. We would like to thank the helpful comments of the referees, in particular for bringing to our attention a number of relevant bibliographic references.


\appendix
\section{Interpolation of supremum}\label{sect:interp}
Polynomial interpolation is the \textit{de facto} standard for problems of this type where the data is regularly spaced and the region of interest is local. In particular spline interpolants, piecewise polynomials constructed to maintain continuity of derivatives, are known to be able to reconstruct a function with high accuracy while using lower order polynomials. The primary advantage of this is to reduce the required support of the contributing data. A number of examples exist for spline interpolants, we will focus on the cubic Hermite spline and a variant of the cubic B-spline. There is a considerable body of literature on spline interpolation, we refer the reader to the texts of \citep{Knott:2000db} and \citep{deBoor:1978jn}. The cubic Hermite spline over an interval of uniform data can be computed for $N$ data points by solving an $N\times N$ tridiagonal system for the slopes at the knots. This yields the following expressions for the interpolating polynomial at the interval $k$ when considering 4 and 6 points respectively:
\begin{align}
 P_{4,k}(s) = \frac{y_{k-1}}{6}s(1-s)(s-2) +\frac{y_k}{2}(1-s^2)(2-s) & +\frac{y_{k+1}}{2}s(2-s)(s+1) \nonumber\\
 & +\frac{y_{k+2}}{6}s(s^2-1), \label{eq:P4}
 \end{align}
 \begin{align}
P_{6,k}(s) = &\frac{y_{k-2}}{90}(7s-12s^2+5s^3)+\frac{4y_{k-1}}{45}(-7s+12s^2-5s^3)\nonumber\\+&\frac{y_{k}}{90}(90-11s-174s^2+95s^3)
+\frac{y_{k+1}}{90}(74s+111s^2-114s^3)\nonumber\\+&\frac{4y_{k+2}}{45}(-2s-3s^2+5s^3)+\frac{y_{k+3}}{90}(2s+3s^2-6s^3),\label{eq:P4}
\end{align}
where the $y_k$ are the values on the collocation points, and $s=\frac{x_p-x_k}{dx}$ is the distance of the point of interest, $x_p$ from the collocation point $x_k$ divided by the spacing, $dx$. One can also perform the global $N\times N$ problem,i.e. $P_{N,k}$, allowing each collocation point to contribute, however this incurs additional computational expense and the influence of the far away points is minimal.

The second interpolation kernel is a variant of the cubic B-spline, developed by \citep{Monaghan1985253} for use in smoothed particle hydrodynamics. The order of accuracy is increased via Richardson extrapolation and it has become a popular method in SPH and vortex methods \citep{cottet,Koumoutsakos:2005p3146,Cottet:1999p20}. The four point interpolant is

\begin{align}
 M_{4,k}'(s) = \frac{y_{k-1}}{2}(-s)(1-s)^2 +\frac{y_k}{2}(2-5s+3s^2) & +\frac{y_{k+1}}{2}(s-4s^2-3s^3) \nonumber\\
 & +\frac{y_{k+2}}{2}s^2(s-1)\label{eq:M4}
\end{align}

Notice these are 1D kernels, their 2D counterparts (bicubic splines) are simply their convolution in each direction. This leads to a computationally simple algorithm; weights for each collocation point in each direction are simply computed as above and combined to form the full interpolant.

Given these choices for cubic splines for computing the value of a variable at a given point, a difficulty still remains as the position of the maximum is unknown and must be located (accurately) as part of the solution. Maximising the bicubic representation is impractical analytically, computationally a more efficient and reliable strategy is a numerical approach, either a Newton method or steepest ascent given a starting point near the collocation point. Unfortunately the robustness of such algorithms is still problematic here, especially where the profile is steepening and values approaching infinity. For instance given a maximum collocation point we will not know which of the four adjacent cells contains the true maximum which will result in four attempts solving the maximisation problem, 3 of which are likely to diverge. A short test was carried out to solve the Newton method for the roots of the derivatives of $P_{4,k}$ and converge on the maximum, results are surprisingly poor compared to the other strategies attempted.

\begin{figure}
\centering
\setlength{\unitlength}{1pt}
\begin{picture}(100,100)
\put(10,10){\circle{3}}
\put(10,50){\circle{3}}
\put(50,10){\circle{3}}
\put(90,90){\circle{3}}
\put(90,50){\circle{3}}
\put(50,90){\circle{3}}
\put(10,90){\circle{3}}
\put(90,10){\circle{3}}
\multiput(30,30)(20,0){3}{\circle*{3}}
\multiput(30,50)(20,0){3}{\circle*{3}}
\multiput(30,70)(20,0){3}{\circle*{3}}
\multiput(20,20)(10,0){7}{\circle*{2}}
\multiput(20,30)(10,0){7}{\circle*{2}}
\multiput(20,40)(10,0){7}{\circle*{2}}
\multiput(20,50)(10,0){7}{\circle*{2}}
\multiput(20,60)(10,0){7}{\circle*{2}}
\multiput(20,70)(10,0){7}{\circle*{2}}
\multiput(20,80)(10,0){7}{\circle*{2}}
\end{picture}	
\caption{\label{sketch} Schematic of point stencil for bisection and \emph{quarter-section} interpolation centred about the current maximum. Unfilled circles are the adjacent collocation points (at the first iteration, neighbouring interpolated points at following steps), larger filled circles the mid point stencil, small filled circles (plus the midpoints) the $7\times7$ interpolated grid.}
\end{figure}
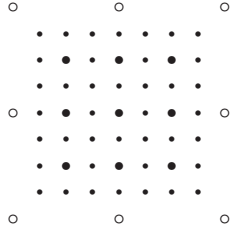

The most straightforward robust approach is to populate the grid cells adjacent to the collocation maximum with a refined grid of interpolated points, and pick from them a new maximum. However this will entail a large number of interpolations and accuracy is limited to the level of refinement chosen. A more efficient method is to develop a 2D `bisection' method, whereby interpolation points are included on a grid at midpoints (i.e. 8 points surrounding the collocation maximum) and from them a new maximum is found (or collocation maximum is retained if it is still largest) and a new set of midpoints (now spaced at $dx/4$ intervals) is populated about the new point. In practise we discovered that in fact a \emph{quarter-section} method outperforms a bisection; at each iteration we populate the $7\times7$ grid of $dx/4$ spaced points, giving a slightly broader support at each step (see figure \ref{sketch}). Note we always use the collocation data points for the interpolation onto the new points, the iterative method is simply an efficient strategy for placing points to search for the maximum.

Note one may also consider employing a more intuitive method whereby a particular function is fit through the collocation points about the collocation maximum and an interpolated maximum extracted from this local representation. An attempt was made to this end using an elliptical gaussian profile and a Newton method to converge on the fit parameters (see supplementary material). The method is appealing as the maximum and its position are immediately given as fit parameters. Accuracy and efficiency could be equivalent to the polynomial methods outlined above, however at late times where the profile steepens the Jacobian matrix of the Newton method becomes ill-conditioned and the method fails. Several work-arounds were explored but none resulted in the accuracy and robustness of the polynomial interpolation where matrix inversions are not required and interpolation weights are readily expressible \textit{a priori} (equations \ref{eq:P4}=\ref{eq:M4}).

\begin{figure}
\centering
\input{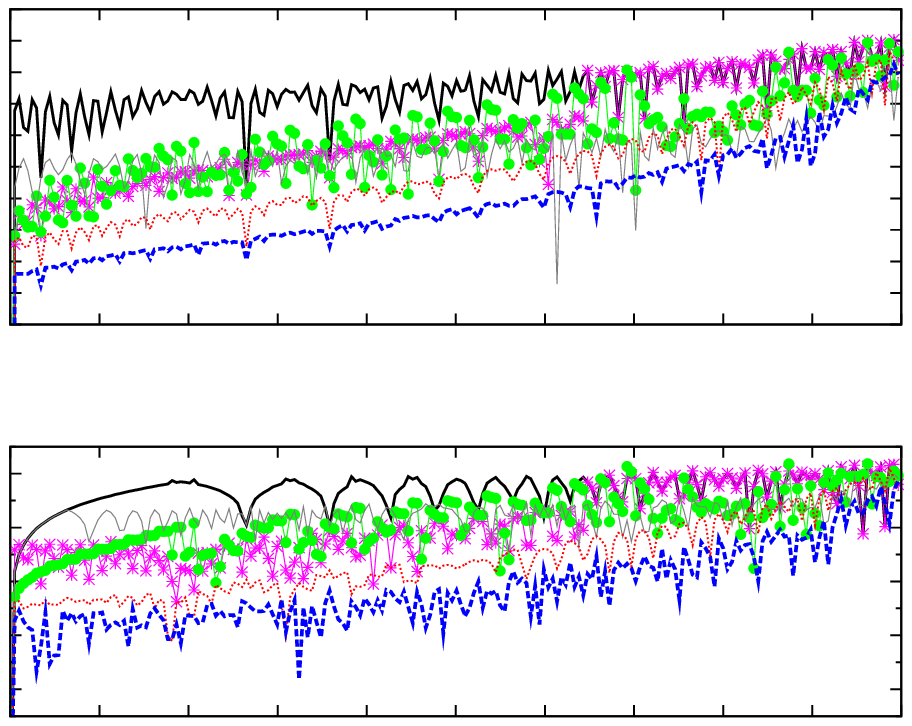}
\caption{\label{fig:interp} Plot of errors for a selection of interpolation methods investigated at resolution $N=512$. Black thick line, no interpolation; blue thick dashed line, $P_{6,k}$ quarter-section; thin red dashed line, $P_{4,k}$ quarter-section; thin grey line, $P_{6,k}$ on $10^2$ points; green circles, $P_{4,k}$ maximisation; magenta stars, Gaussian fit.}
\end{figure}

To assess the performance of each method we compute the errors $\mathcal{E}_\gamma(t)$ and $\mathcal{E}_\omega(t)$ as defined in section \ref{sec:model} over a simulation of the $\lambda=3/2$ model system (unmapped) at $512^2$ resolution for $T=1.0$ (see reliability time estimate in table \ref{tab:Reliability}). In addition to the error measures, we determine an estimate of the average CPU time by computing 1000 executions of the interpolation algorithm. Table \ref{table:interp} shows the errors of each interpolation method and CPU times. It was found that the accumulation of numerical error at late times, i.e. as the singularity time is approached,  dominates the integral measures so we show the error measures of a truncated time series ($T=0.8$) which gives a more reliable measure of the errors throughout the simulation (see figure \ref{fig:interp}).

\begin{table}
\centering
    \begin{tabular}{ l  c  c  c c c}
    Interpolation & CPU (s) & $Q_\gamma(T=1.0)$ & $Q_\omega(T=1.0)$ & $\frac{\|\mathcal{E}_\gamma\|_2}{T}(T=1.0)$&$\frac{\| \mathcal{E}_\omega\|_2}{T}(T=1.0)$\\ 
 \\
None & 7.8 & 6.2E-8 & 6.0E-5 & 0.19 & 1.32 \\
$P_{N,k}$ quarter-section & 374 & 2.1E-10 & 7.1E-7 & 0.03 & 0.19 \\
$P_{6,k}$ quarter-section &17.1 & 6.7E-10 & 1.1E-6 & 0.038 & 0.20\\
$P_{4,k}$ quarter-section & 16.7 & 5.1E-9 & 1.2E-5 & 0.07 & 0.44 \\
$P_{N,k}$ on $100^2$ points & 74.5 & 2.7E-10 & 3.4E-6 & 0.032 & 0.30\\
$P_{N,k}$ on $10^2$ points & 72.0 & 4.0E-10 & 4.4E-6 & 0.038 & 0.49 \\
$P_{6,k}$ on $100^2$ points & 71.2 & 4.8E-10 & 4.0E-6 & 0.037 & 0.31 \\
$P_{6,k}$ on $10^2$ points & 16.8 & 6.0E-10 & 6.5E-6 & 0.039 & 0.51 \\
$P_{4,k}$ on $100^2$ points & 43.4 & 5.4E-10 & 1.3E-5 & 0.071 & 0.46 \\
$P_{4,k}$ on $10^2$ points & 16.3 & 5.9E-10 & 1.4E-5 & 0.074 & 0.57 \\
$\mathrm{M'}_{4,k}$ on $100^2$ points &34.2 &6.9E-9 & 1.9E-5 & 0.078 & 0.53 \\
$\mathrm{M'}_{4,k}$ on $10^2$ points &16.4 &7.0E-9 & 2.0E-5 & 0.078 & 0.62 \\
Gaussian fit&16.9 & 6.2E-8 & 6.0E-5 & 0.18 & 1.0 \\
$P_{4,k}$ maximisation&16.4 & 2.1E-8 & 2.9E-5 & 0.11& 0.77 \\
    \end{tabular}

    \begin{tabular}{ l  c  c  c c c}
    \hline
    Interpolation & CPU (s)  & $Q_\gamma(T=0.8)$ & $Q_\omega(T=0.8)$ & $\frac{\|\mathcal{E}_\gamma\|_2}{T}(T=0.8)$&$\frac{\| \mathcal{E}_\omega\|_2}{T}(T=0.8)$\\ 
 \\
None & 7.8 & 2.8E-9 & 1.5E-5 & 0.14 & 1.28 \\
$P_{N,k}$ quarter-section & 374 & 7.3E-15 & 8.8E-11 & 0.0034 & 0.036 \\
$P_{6,k}$ quarter-section &17.1 & 4.4E-15 & 5.2E-11 & 0.0029 & 0.028 \\
$P_{4,k}$ quarter-section & 16.7  & 5.1E-13 & 5.1E-9 & 0.011 & 0.097 \\
$P_{N,k}$ on $100^2$ points & 74.5 &  7.3E-15 & 1.3E-9 & 0.0037 & 0.13 \\
$P_{N,k}$ on $10^2$ points & 72.0 &  4.1E-13 & 1.5E-7 & 0.014 & 0.42 \\
$P_{6,k}$ on $100^2$ points & 71.2 &  3.4E-15 & 1.25E-9 & 0.0028 & 0.13 \\
$P_{6,k}$ on $10^2$ points & 16.8 &  3.1E-13 & 1.5E-7 & 0.0135 & 0.42 \\
$P_{4,k}$ on $100^2$ points & 43.4 & 5.3E-13 & 7.6E-9 & 0.011 & 0.15\\
$P_{4,k}$ on $10^2$ points & 16.3 &  1.38E-12 & 1.5E-7 & 0.017 & 0.42 \\
$\mathrm{M'}_{4,k}$ on $100^2$ points &34.2 & 1.3E-12 & 8.4E-8 & 0.015 & 0.25 \\
$\mathrm{M'}_{4,k}$ on $10^2$ points &16.4 & 2.1E-12 & 2.9E-7 & 0.018 & 0.44 \\
Gaussian fit&16.9 &  2.3E-9 & 1.2E-5 & 0.094 & 0.73 \\
$P_{4,k}$ maximisation&16.4 & 1.3E-10 & 5.0E-6 & 0.042 & 0.59
    \end{tabular}
       \caption{Table showing CPU and error measures for the interpolation methods investigated at resolution $N=512$ and $T=1.0$(top) and $T=0.8$(bottom). Note CPU time is evaluated by 1000 executions of the interpolation subroutine. No interpolation simply means we take the maximal collocation value. Gaussian fails after $t=0.65$ and reverts to collocation points thereafter. $\omega_{\mathrm{num}}(X_+(t),Y_+(t),t)$ is evaluated in each case using $P_{N,k}$ but $(X_+(t),Y_+(t),t)$ found from the method indicated.}
    \label{table:interp}
\end{table}

Figure \ref{fig:interp} and table \ref{table:interp} show that errors are consistently smallest for the $P_{N,k}$ and $P_{6,k}$ interpolants, better for the cases where $100\times 1000$ points are used or the quarter-section method. Considering the computational cost of each method it becomes quite clear that the most efficient method is the $P_{6,k}$ quarter-section. The quarter-section method will entail a far fewer interpolations to reach a similar accuracy than, for example the $100\times 1000$ equivalent. We find the Gaussian fit to be poor, primarily due to it failing a $t=0.65$ (see figure \ref{fig:interp}). The $P_{4,k}$ maximisation (root finding) is also found to be inaccurate compared to the `point-search' methods. We suspect this to be where the Newton method converges (reaches machine precision tolerance in the residual) at a point which is not the true maximum. It might be possible to tune this method by using a coarse sweep (i.e. the first quarter-section grid) before commencing a Newton solve from a closer initial guess, however the quarter-section method proves so fast and accurate that for our purposes we will retain it as our default interpolation.


\bibliographystyle{jfm}
\bibliography{papers}

\begin{thebibliography}{43}
\expandafter\ifx\csname natexlab\endcsname\relax\def\natexlab#1{#1}\fi

\bibitem[Ayala \& Protas(2014)]{Ayala:2014hi}
{\sc Ayala, Diego \& Protas, Bartosz} 2014 {Maximum palinstrophy growth in 2D
  incompressible flows}. {\em Journal of Fluid Mechanics\/} {\bf 742},
  340--367.

\bibitem[Bardos \& Titi(2007)]{Bardos07eulerequations}
{\sc Bardos, Claude \& Titi, Edriss~S.} 2007 Euler equations of incompressible
  ideal fluids.

\bibitem[Beale {\em et~al.\/}(1984)Beale, Kato \& Majda]{BKM84}
{\sc Beale, J.~T., Kato, T. \& Majda, A.} 1984 Remarks on the breakdown of
  smooth solutions for the 3-d euler equations. {\em Communications in
  Mathematical Physics\/} {\bf 94}, 61--66.

\bibitem[de~Boor(1978)]{deBoor:1978jn}
{\sc de~Boor, Carl} 1978 {A Practical Guide to Splines}. Springer New York, New
  York, NY.

\bibitem[Brachet {\em et~al.\/}(2013)Brachet, Bustamante, Krstulovic, Mininni,
  Pouquet \& Rosenberg]{brachet2013ideal}
{\sc Brachet, ME, Bustamante, MD, Krstulovic, G, Mininni, PD, Pouquet, A \&
  Rosenberg, D} 2013 Ideal evolution of magnetohydrodynamic turbulence when
  imposing taylor-green symmetries. {\em Physical Review E\/} {\bf 87}~(1),
  013110.

\bibitem[Bustamante(2011)]{bustamante20113d}
{\sc Bustamante, Miguel~D} 2011 {3D Euler equations and ideal MHD mapped to
  regular systems: Probing the finite-time blowup hypothesis}. {\em Physica D:
  Nonlinear Phenomena\/} {\bf 240}~(13), 1092--1099.

\bibitem[Bustamante \& Brachet(2012)]{bustamante2012interplay}
{\sc Bustamante, Miguel~D \& Brachet, Marc} 2012 Interplay between the
  beale-kato-majda theorem and the analyticity-strip method to investigate
  numerically the incompressible euler singularity problem. {\em Physical
  Review E\/} {\bf 86}~(6), 066302.

\bibitem[Bustamante \& Kerr(2008)]{Bustamante:2008p2289}
{\sc Bustamante, Miguel~D \& Kerr, Robert~M} 2008 {3D Euler about a 2D symmetry
  plane}. {\em Physica D: Nonlinear Phenomena\/} {\bf 237}~(14-17), 1912--1920.

\bibitem[Cantwell(1992)]{Cantwell:1992jp}
{\sc Cantwell, Brian~J} 1992 {Exact solution of a restricted Euler equation for
  the velocity gradient tensor}. {\em Phys. Fluids A\/} {\bf 4}~(4), 782--793.

\bibitem[Constantin(2000)]{Constantin:2000fa}
{\sc Constantin, Peter} 2000 {The Euler equations and nonlocal conservative
  Riccati equations}. {\em International Mathematics Research Notices\/} ~(9),
  455--465.

\bibitem[Cottet {\em et~al.\/}(1999)Cottet, Salihi \&
  El~Hamraoui]{Cottet:1999p20}
{\sc Cottet, GH, Salihi, MLO \& El~Hamraoui, M} 1999 {Multi-purpose regridding
  in vortex methods}. {\em ESAIM Proceedings\/} pp. 94--103.

\bibitem[Cottet \& Koumoutsakos(2000)]{cottet}
{\sc Cottet, Georges-Henri \& Koumoutsakos, Petros~D} 2000 {\em {Vortex
  methods}\/}. Cambridge Univ Pr.

\bibitem[Deng {\em et~al.\/}(2005)Deng, Hou \& Yu]{Deng:2005p2138}
{\sc Deng, Jian, Hou, Thomas~Y \& Yu, Xinwei} 2005 {Geometric Properties and
  Nonblowup of 3D Incompressible Euler Flow}. {\em Communications in Partial
  Differential Equations\/} {\bf 30}~(1--2), 225--243.

\bibitem[Donzis {\em et~al.\/}(2013)Donzis, Gibbon, Gupta, Kerr, Pandit \&
  Vincenzi]{Donzis:2013ed}
{\sc Donzis, Diego~A, Gibbon, John~D, Gupta, Anupam, Kerr, Robert~M, Pandit,
  Rahul \& Vincenzi, Dario} 2013 {Vorticity moments in four numerical
  simulations of the 3D Navier-Stokes equations}. {\em Journal of Fluid
  Mechanics\/} {\bf 732}, 316--331.

\bibitem[Euler(1761)]{Euler1752}
{\sc Euler, L.} 1761 Principia motus fluidorum. {\em Novi Commentarii Acad.
  Sci.Petropolitanae\/} {\bf 6}, 271--311.

\bibitem[Frisch \& Villone(2014)]{Frisch:2014gl}
{\sc Frisch, Uriel \& Villone, Barbara} 2014 {Cauchy{\textquoteright}s almost
  forgotten Lagrangian formulation of the Euler equation for 3D incompressible
  flow}. {\em The European Physical Journal H\/} {\bf 39}~(3), 325--351.

\bibitem[Gibbon(2008)]{Gibbon20081894}
{\sc Gibbon, J.D.} 2008 The three-dimensional {E}uler equations: {W}here do we
  stand? {\em Physica D: Nonlinear Phenomena\/} {\bf 237}, 1894 -- 1904, euler
  Equations: 250 Years On, Proceedings of an international conference.

\bibitem[Gibbon {\em et~al.\/}(1999)Gibbon, Fokas \& Doering]{Gibbon1999497}
{\sc Gibbon, J.D., Fokas, A.S. \& Doering, C.R.} 1999 Dynamically stretched
  vortices as solutions of the 3d navier-stokes equations. {\em Physica D:
  Nonlinear Phenomena\/} {\bf 132}~(4), 497 -- 510.

\bibitem[Gibbon(2013)]{Gibbon:2013cg}
{\sc Gibbon, J~D} 2013 {Dynamics of Scaled Norms of Vorticity for the
  Three-dimensional Navier-Stokes and Euler Equations}. {\em Procedia IUTAM\/}
  {\bf 7}, 39--48.

\bibitem[Gibbon {\em et~al.\/}(2003)Gibbon, Moore \& Stuart]{Gibbon:2003ix}
{\sc Gibbon, J~D, Moore, D~R \& Stuart, J~T} 2003 {Exact, infinite energy,
  blow-up solutions of the three-dimensional Euler equations}. {\em
  Nonlinearity\/} {\bf 16}~(5), 1823--1831.

\bibitem[Gibbon \& Ohkitani(2001)]{0951-7715-14-5-316}
{\sc Gibbon, J~D \& Ohkitani, K} 2001 Singularity formation in a class of
  stretched solutions of the equations for ideal magneto-hydrodynamics. {\em
  Nonlinearity\/} {\bf 14}~(5), 1239.

\bibitem[Grafke \& Grauer(2013)]{Grafke:2013dy}
{\sc Grafke, Tobias \& Grauer, Rainer} 2013 {Finite-time Euler singularities: A
  Lagrangian perspective}. {\em Applied Mathematics Letters\/} {\bf 26}~(4),
  500--505.

\bibitem[Grafke {\em et~al.\/}(2008)Grafke, Homann, Dreher \&
  Grauer]{grafke2008numerical}
{\sc Grafke, Tobias, Homann, Holger, Dreher, J{\"u}rgen \& Grauer, Rainer} 2008
  Numerical simulations of possible finite time singularities in the
  incompressible euler equations: comparison of numerical methods. {\em Physica
  D: Nonlinear Phenomena\/} {\bf 237}~(14), 1932--1936.

\bibitem[Hou \& Li(2006)]{Hou:2006p2135}
{\sc Hou, Thomas~Y \& Li, Ruo} 2006 {Dynamic Depletion of Vortex Stretching and
  Non-Blowup of the 3-D Incompressible Euler Equations}. {\em Journal of
  Nonlinear Science\/} {\bf 16}~(6), 639--664.

\bibitem[Hou \& Li(2007)]{Hou:2007}
{\sc Hou, Thomas~Y \& Li, Ruo} 2007 {Computing nearly singular solutions using
  pseudo-spectral methods}. {\em Journal of Computational Physics\/} {\bf
  226}~(1), 379--397.

\bibitem[Kerr(1993)]{Kerr:1993p2227}
{\sc Kerr, RM} 1993 {Evidence for a singularity of the three-dimensional,
  incompressible Euler equations}. {\em Physics of Fluids A Fluid Dynamics\/}
  {\bf 5}, 1725.

\bibitem[Kerr(2013)]{Kerr:2013gh}
{\sc Kerr, Robert~M} 2013 {Bounds for Euler from vorticity moments and line
  divergence}. {\em Journal of Fluid Mechanics\/} {\bf 729}, R2.

\bibitem[Kiselev \& Zlatos(2014)]{Kiselev:2014ti}
{\sc Kiselev, Alexander \& Zlatos, Andrej} 2014 {Blow up for the 2D Euler
  Equation on Some Bounded Domains}. {\em arXiv.org\/} ~(1406.3648v1), math.AP.

\bibitem[Knott(2000)]{Knott:2000db}
{\sc Knott, Gary~D} 2000 {Interpolating Cubic Splines}. Birkh{\"a}user Boston,
  Boston, MA.

\bibitem[{Kolmogorov}(1941)]{1941DoSSR..30..301K}
{\sc {Kolmogorov}, A.} 1941 {The Local Structure of Turbulence in
  Incompressible Viscous Fluid for Very Large Reynolds' Numbers}. {\em
  Akademiia Nauk SSSR Doklady\/} {\bf 30}, 301--305.

\bibitem[Koumoutsakos(2005)]{Koumoutsakos:2005p3146}
{\sc Koumoutsakos, P} 2005 {Multiscale flow simulations using particles}. {\em
  Annual Reviews\/} {\bf 37}, 457--87.

\bibitem[Kuznetsov(2006)]{kuznetsov2006vortex}
{\sc Kuznetsov, EA} 2006 Vortex line representation for the hydrodynamic type
  equations. {\em Journal of Nonlinear Mathematical Physics\/} {\bf 13}~(1),
  64--80.

\bibitem[Kuznetsov \& Ruban(2000)]{kuznetsov2000hamiltonian}
{\sc Kuznetsov, EA \& Ruban, VP} 2000 Hamiltonian dynamics of vortex and
  magnetic lines in hydrodynamic type systems. {\em Physical Review E\/} {\bf
  61}~(1), 831.

\bibitem[Li \& Rodrigo(2009)]{Rodrigo2009}
{\sc Li, Dong \& Rodrigo, Jose} 2009 Blow up for the generalized surface
  quasi-geostrophic equation with supercritical dissipation. {\em
  Communications in Mathematical Physics\/} {\bf 286}~(1), 111--124.

\bibitem[Lu \& Doering(2008)]{Lu:2008dz}
{\sc Lu, Lu \& Doering, Charles~R} 2008 {Limits on enstrophy growth for
  solutions of the three-dimensional Navier-Stokes equations}. {\em Indiana
  University Mathematics Journal\/} {\bf 57}~(6), 2693--2728.

\bibitem[Luo \& Hou(2014)]{Luo:2014jg}
{\sc Luo, G \& Hou, T~Y} 2014 {Potentially singular solutions of the 3D
  axisymmetric Euler equations}. {\em Proceedings of the National Academy of
  Sciences\/} {\bf 111}~(36), 12968--12973.

\bibitem[Mailybaev(2013)]{PhysRevE.87.053011}
{\sc Mailybaev, Alexei~A.} 2013 Blowup as a driving mechanism of turbulence in
  shell models. {\em Phys. Rev. E\/} {\bf 87}, 053011.

\bibitem[Monaghan(1985)]{Monaghan1985253}
{\sc Monaghan, JJ} 1985 {Extrapolating B splines for interpolation}. {\em
  Journal of Computational Physics\/} {\bf 60}~(2), 253--262.

\bibitem[Ohkitani \& Gibbon(2000)]{Ohkitani20003181}
{\sc Ohkitani, K. \& Gibbon, J.D.} 2000 Numerical study of singularity
  formation in a class of euler and navier-stokes flows. {\em Physics of
  Fluids\/} {\bf 12}~(12), 3181--3194, cited By (since 1996)35.

\bibitem[Orlandi {\em et~al.\/}(2014)Orlandi, Pirozzoli, Bernardini \&
  Carnevale]{Orlandi:2014kl}
{\sc Orlandi, Paolo, Pirozzoli, Sergio, Bernardini, Matteo \& Carnevale,
  George~F} 2014 {A minimal flow unit for the study of turbulence with passive
  scalars}. {\em dx.doi.org\/} {\bf 15}~(11), 731--751.

\bibitem[Perlin \& Bustamante(2014)]{Perlin2014}
{\sc Perlin, Marc \& Bustamante, Miguel~D.} 2014 {A Robust Quantitative
  Comparison Criterion of Two Signals based on the Sobolev Norm of Their
  Difference}. {\em arXiv.org\/} ~(1412.6977), physics.flu--dyn.

\bibitem[Sulem {\em et~al.\/}(1983)Sulem, Sulem \& Frisch]{Sulem:1983}
{\sc Sulem, C, Sulem, P~L \& Frisch, H} 1983 {Tracing complex singularities
  with spectral methods}. {\em Journal of Computational Physics\/} {\bf
  50}~(1), 138 -- 161.

\bibitem[Vieillefosse(1984)]{1984PhyA..125..150V}
{\sc Vieillefosse, P} 1984 {Internal Motion of a Small Element of Fluid in an
  Inviscid Flow}. {\em Physica A\/} {\bf 125}~(1), 150--162.

\end{thebibliography}

\end{document}